\documentclass[useASM,usenatbib]{mn2e}

\usepackage{graphicx}
\usepackage{multirow}
\usepackage{amssymb}
\usepackage{hyperref}

\title[SN 2009bw in UGC2890]{The bright Type IIP SN 2009bw, showing signs of interaction
\thanks{Based on observations collected at the Italian 3.58m Telescopio Nazionale Galileo, the Liverpool Telescope, the Nordic Optical Telescope (La Palma, Spain),  the Calar Alto 2.2m Telescope (Sierra de los Filabres, Spain), the orbital telescope SWIFT, the Copernico and Galileo Galilei telescopes (Asiago, Italy), the Special Astrophysical Observatory (Mt. Pastukhov, Russia), the Taurus Hill Observatory (Hill H\"ark\"am\"aki, Finland), the 0.5m Telescope (Tatranska Lomnica, Slovakia), the 0.5m and 0.6m telescopes of Sternberg Atronomical Institute Observatory (Nauchnyi, Crimea) and the 0.7m Telescope (Moscow, Russia). } }
\author[C. Inserra et al.]{C. Inserra$^{1}$$^{,}$$^{2}$$^{,}$$^{3}$\thanks{E-mail: cosimo.inserra@oact.inaf.it(CI)}, M. Turatto$^{4}$, A. Pastorello$^{5,6}$, M.L. Pumo$^{6,7}$, E. Baron$^{3,8}$, S. Benetti$^{6}$,
\newauthor E. Cappellaro$^{6}$, S.Taubenberger$^{9}$, F. Bufano$^{2,6}$,  N. Elias-Rosa$^{10}$, L. Zampieri$^{6}$,
\newauthor  A. Harutyunyan$^{11}$, A. S. Moskvitin$^{12}$, M. Nissinen$^{13}$, V. Stanishev$^{14}$, D. Y. Tsvetkov$^{15}$,
\newauthor V.P. Hentunen$^{13}$, V.N. Komarova$^{12}$, N.N. Pavlyuk$^{15}$, V.V. Sokolov$^{12}$, T.N. Sokolova$^{12}$.\\
\\
$^{1}$Dipartimento di Fisica ed Astronomia, Universita' di Catania, Sezione Astrofisica, Via S.Sofia 78, 95123, Catania, Italy\\
$^{2}$INAF Osservatorio Astrofisico di Catania, Via S.Sofia 78, 95123, Catania, Italy\\
$^{3}$Department of Physics and Astronomy, University of Oklahoma, Norman, OK 73019\\
$^{4}$INAF Osservatorio Astronomico di Trieste, Via Tiepolo 11, 34143, Trieste, Italy\\
$^{5}$Astrophysics Research Centre, School of Mathematics and Physics, Queen's University Belfast, Belfast BT7 1NN, United Kingdom\\
$^{6}$INAF Osservatorio Astronomico di Padova, Vicolo dell'Osservatotio 5, 35122, Padova, Italy\\
$^{7}$Dipartimento di Astronomia, Universita' di Padova, Vicolo dell'Osservatorio 3, 35122, Padova, Italy\\ 
$^{8}$Hamburger Sternwarte, Gojenbergsweg 112, 21029 Hamburg, Germany\\
$^{9}$Max-Planck-Institut f\"ur Astrophysik, Karl-Schwarzschild-Str. 1, 85741 Garching, Germany\\
$^{10}$Institut de Cincies de l'Espai (IEEC-CSIC), Campus UAB, 08193 Bellaterra, Spain\\
$^{11}$Fundaci\'on Galileo Galilei-INAF, Telescopio Nazionale Galileo, Rambla Jos\'e Ana Fern\'andez P\'erez 7, 38712 Bre\~na Baja, TF - Spain\\
$^{12}$Special Astrophysical Observatory of the Russian AS, Nizhnij Arkhyz 369167, Russia\\
$^{13}$Taurus Hill Observatory, Kangaslampi, Finland\\
$^{14}$CENTRA - Instituto Superior Tecnico, Av. Rovisco Pais, 1,1049-001 Lisbon, Portogual\\
$^{15}$Sternberg Astronomical Institute of Lomonosov Moscow State University, University Ave., 13, 119992 Moscow, Russia}

\def\kms{km\,s$^{-1}$}
\def\Ha{H{$\alpha$}}
\def\Hb{H{$\beta$}}

\def\ni{$^{56}$Ni}
\def\co{$^{56}$Co}
\def\fe{$^{56}$Fe}

\def\em{SN~1999em}

\def\h{SN~1992H}
\def\od{SN~2007od}
\def\bw{SN~2009bw}

\def\mcento{mag\,(100d)$^{-1}$}

\def\M{M$_{\odot}$}

\def\ebv{E(B--V)}

\begin{document}

\date{Received.....; accepted...........}

\pagerange{\pageref{firstpage}--\pageref{lastpage}} \pubyear{}

\maketitle

\label{firstpage}

\begin{abstract}
We present photometry and spectroscopy of the type IIP supernova 2009bw in UGC 2890 from few days after the outburst to 241 days. The light curve of \bw\/ during the photospheric phase is similar to that of normal SNe IIP but with brighter peak and plateau (M$_{R}^{max}= -17.82$ mag, M$_{R}^{plateau}= -17.37$ mag). The luminosity drop from the photospheric to the nebular phase is one of the fastest ever observed, $\sim$2.2 mag in about 13 days. The radioactive tail of the bolometric light curve indicates that the amount of ejected \ni\/ is $\approx$ 0.022 \M\/. The photospheric spectra reveal high velocity lines of \Ha\/ and \Hb\/ until about 105 days after the shock breakout, suggesting a possible early interaction between the SN ejecta and pre-existent circumstellar material, and the presence of CNO elements. By modeling the bolometric light curve, ejecta expansion velocity and photospheric temperature, we estimate a total ejected mass of $\sim8-12$\M\/, a kinetic energy of $\sim$0.3 foe and an initial radius of $\sim3.6-7\times10^{13}$ cm.
 \end{abstract}


\section{Introduction}\label{sec:intro}

Type II Supernovae (SNe) are produced by the explosion following the gravitational collapse of massive stars \citep[M$_{ZAMS}$$\gtrsim7-8 M$$_{\odot}$, for details see][and reference therein]{pumo09,smart,he03}.

Those retaining part of their H envelope at the time of the explosion are called type II and, if they show a constant luminosity for a period ranging from about 30 days to a few months, are assigned to the \textquotedblleft Plateau" subclass (SNe IIP). Others present a steeper decline over the same period and are named \textquotedblleft Linear" \citep[SNe IIL;][]{barbon}. A third sub-class shows a slow decline and narrow emission lines in the spectra, and hence forms the so called group of \textquotedblleft narrow" type II SNe \citep[SNe IIn,][]{IIn}. SNe IIP have been the subject of extensive analysis by \citet{hamuy} who pointed out a continuity in their properties and revealed several relations linking the observables to physical parameters. An independent analysis, extending the sample to low- and high-luminosity SNe IIP confirmed these results \citep{pphdt} . 
Thanks to the direct identification of several SN precursors in deep pre-explosion images, recent studies support the idea that most SNe IIP are originated from explosions of stars with M $\lesssim$ 16-17 M$_{\odot}$ \citep{smart} and are usually associated with red supergiant progenitors \citep{smart1}.
Such stars have \textquotedblleft extended", massive ($\gtrsim5-7$ \M\/) hydrogen envelopes at the time of explosion. Variation in the explosion configurations (e.g. envelope mass, energy, radius, etc.) are thought to be responsible for the relatively \textquotedblleft large" variety of objects.

\begin{figure}
\includegraphics[width=\columnwidth]{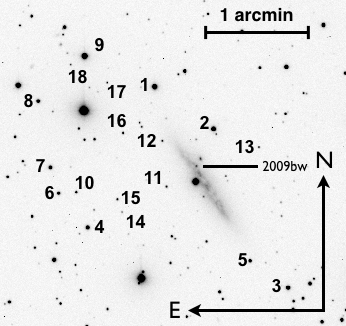}
\caption{R band image of SN 2009bw in UGC 2890 obtained with the Calar Alto 2.2m Telescope+CAFOS (see Tab.~\ref{table:cht}) on August 30th, 2009. The sequence of stars used to calibrate the optical and NIR magnitudes of SN 2009bw is indicated.} 
\label{fig:09bw}
\end{figure}

 
The configuration of the exploding stars and of the circumstellar matter (CSM), governed by the mass loss during the late evolutionary stages of the progenitor, can lead to different observables starting from stars of similar initial mass. Detailed studies at many different wavelengths show an almost unlimited variety of interaction scenarios between ejecta of different mass and CSM of various densities and distances from the exploding stars. As a recent example, \od\/ \citep{07od} shows evidence of weak interaction at early time and strong interaction at late times.

\bw\/ is a good opportunity to explore the zoo of type IIP with early, weak interaction. 
 It was discovered in UGC 2890 on 2009 March 27.87 UT \citep{c1}. \citet{c2} classified it as a normal type II SN soon after explosion, showing a spectrum with narrow \Ha\/ emission superimposed on a broader base.

The coordinates of \bw\/ have been measured on our astrometrically calibrated images:
$\alpha$ = 03$^{h}$56$^{m}$06$^{s}$.92 $\pm$0$^{s}$.05 $\delta$ = +72$^{o}$55'40".90 $\pm$0."05 (J2000). The object is located in the inner region of the highly (i=90, LEDA) inclined UGC 2890, 11" East and 22" North of the center of the galaxy (Fig.~\ref{fig:09bw}).
This corresponds to a projected distance of $\sim$2.4 Kpc from the nucleus assuming a distance to UGC 2890 of $\sim$20 Mpc (see Sect.~\ref{sec:red}).


In this paper we present and discuss optical and near-infrared observations of \bw\/ from March 2009 to August 2010.
In Sect.~\ref{sec:phot} we describe the photometric observations and give details on the reduction process, the reddening estimate and the photometric evolution. In Sect.~\ref{sec:spec} we describe and analyze the spectroscopic data. The explosion and progenitor parameters retrieved through the models are presented in Sect.~\ref{sec:m}. Discussion and conclusions follow in Sect.~\ref{sec:dis} and~\ref{sec:final}. 

\section{Photometry}\label{sec:phot}
The optical photometric monitoring of SN 2009bw started on March 30th, 2009, the day after the discovery, and continued until late November 2009. 
We attempted to recover the object at late time after the seasonal gap but the object was already fainter than our instrument detection limit.

\begin{figure*}
\includegraphics[width=18cm]{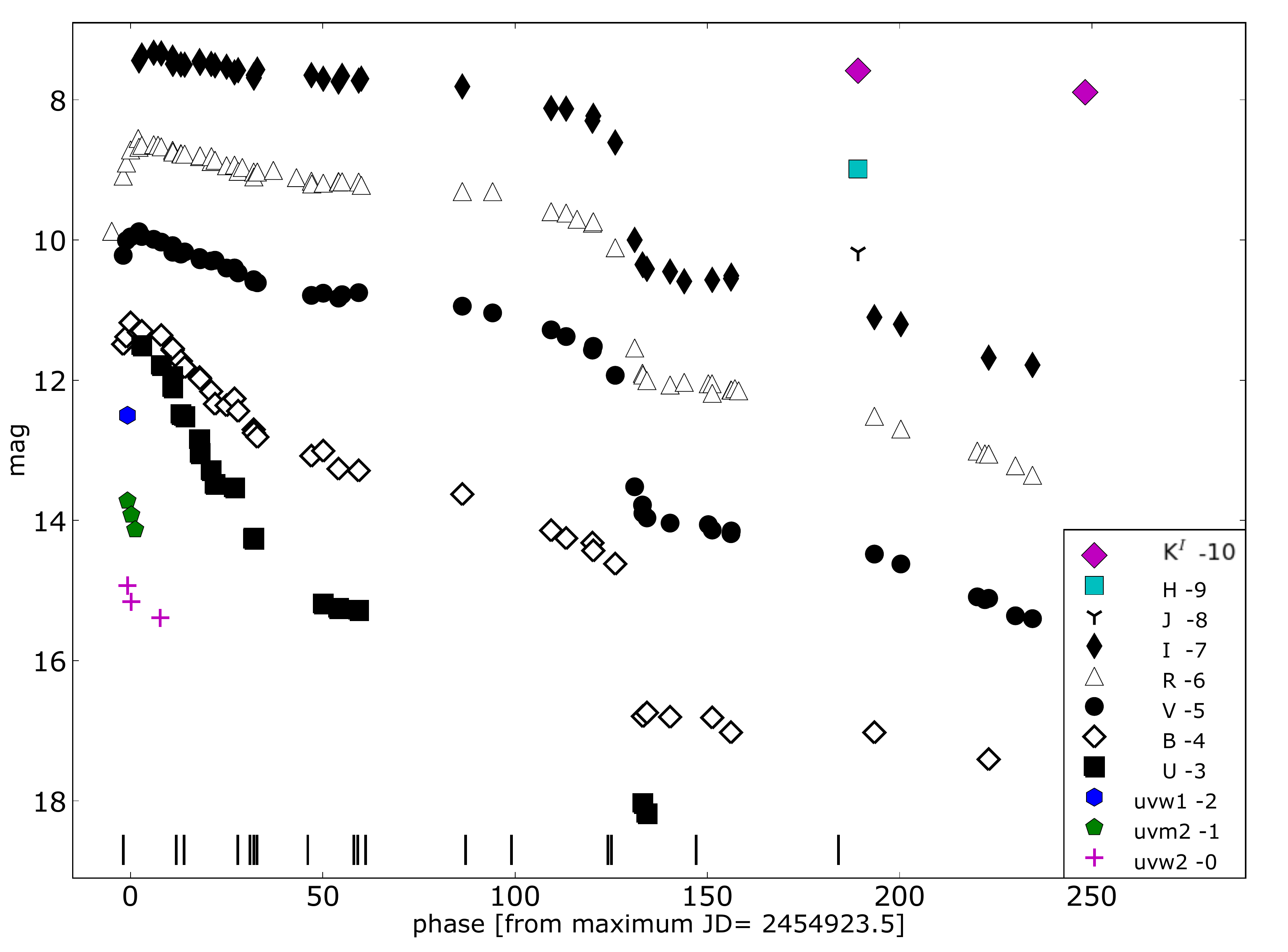}
\caption{Synoptic view of the light curves of SN 2009bw  in all available bands. The shifts from the original values reported on Tab.~\ref{table:snm} are in the legend. Vertical marks at the bottom indicate the epochs of available spectra (cfr. Tab.~\ref{table:sp}).} 
\label{fig:sn_lc}
\end{figure*}

\begin{table*}
\caption{Instrument setup used for the photometric follow up.}
\begin{center}\tabcolsep=1.0mm
\begin{tabular}{lclclccl}
\hline
\hline
Telescope & Primary mirror & Camera & Array & CCD & pixel scale & field of view & filters \\
 & m & & &  & arcsec/pix & arcmin & \\
 \hline
 Copernico & 1.82 & AFOSC & 1024 x 1024 & TK1024AB & 0.46 & 8.1 & Bessell B V R, Gunn i\\
 TNG & 3.58 & DOLORES & 2048 x 2048 & EEV 42-40 & 0.25 & 8.6 & Johnson U B V, Cousins R I\\
 				&  & NICS & 1024 x 1024 & HgCdTe Hawaii & 0.25 & 4.2  & J H K$^{\prime}$\\
 LT & 2.0 & RATcam & 2048 x 2048 &  EEV 42-40 & 0.13 & 4.6 & Sloan u, Bessell B V, Sloan r i\\
 NOT & 2.5 & ALFOSC & 2048 x 2048 & EEV 42-40 & 0.19 & 6.4 & Johnson U B V R, interference i \\
 CAHA & 2.2 & CAFOS & 2048 x 2048 & SITe & 0.53 & 16 & Johnson U B V R I\\
SWIFT& 0.3 & UVOT & 2048 x 2048 &  microchannel & 0.48 & 17 & uvw2,uvm2,uvw1 \\
 & & & &intensified CCD & & &\\
  SAO-RAS & 1.0 & CCD phot. & 530 x 580 & EEV 42-40 & 0.48 & 2.38 x 3.53 &  Johnson U B V, Cousins R I\\
  THO & 0.4 & ST-8XME & 1530 x 1020 & KAF-1603ME & 0.9 & 24 x 16 &  Johnson B V, Cousins R\\
  S50 & 0.5 & ST-10XME & 2184 x 1472 & KAF-3200ME & 1.12 & 20.6 x 13.9 &  Johnson V, Cousins R I\\
  M70 & 0.7 & Apogee AP-7p & 512 x 512 & SITe & 0.64 & 5.5 &  Johnson U B V I, Cousins R\\
  C60 & 0.6 & Apogee AP-47p & 1024 x 1024 & EEV47-10 & 0.71 & 6.1 &  Johnson B V I, Cousins R\\
  C50 & 0.5 & Meade Pictor416 & 765 x 510 & KAF-0400 & 0.92 & 11.7 x 7.8 &  Johnson V, Cousins R\\
 \hline
\end{tabular}
Copernico = Copernico Telescope (Mt. Ekar, Asiago, Italy); TNG = Telescopio Nazionale Galileo (La Palma, Spain); LT = Liverpool Telescope (La Palma, Spain); NOT = Nordic Optical Telescope (La Palma, Spain); CAHA = Calar Alto Observatory 2.2m Telescope (Sierra de los Filabres, Andalucia, Spain); SWIFT orbiting telescope by NASA; SAO-RAS = Special Astrophysical Observatory 1m Telescope (Mt. Pastukhov, Russia); THO = Taurus Hill Observatory (Hill H\"ark\"am\"aki, Finland); S50 = 0.5m Newton Telescope (Tatranska Lomnica, Slovakia); M70 = 0.7m Cassegrain Telescope (Moscow, Russia); C60 = 0.6m Cassegrain Telescope (Observatory of Sternberg Astronomical Institute, Nauchnyi, Crimea); C50 = 0.5m Matsukov Telescope Meniscus (Observatory of SAI, Nauchnyi, Crimea).
\end{center}
\label{table:cht}

\end{table*}%

\subsection{Data}

Ground based optical photometry was obtained with several telescopes (Tab.~\ref{table:cht}).
Note that the u, r and i magnitudes of the Sloan filters of the LT and the $i$-band filter of the NOT have been reported in the Johnson--Cousin system.
Optical data were reduced following standard procedures in the IRAF\footnote{Image Reduction and Analysis Facility, distributed by the National Optical Astronomy Observatories, which are operated by the Association of Universities for Research in Astronomy, Inc, under contract to the National Science Foundation.} environment. Instrumental magnitudes were measured on the final images, obtained after removal of the detector signature including overscan correction, bias subtraction, flat field correction and trimming.

\begin{table}
  \caption{Magnitudes of the local sequence stars in the field of SN 2009bw (cfr. Fig.~\ref{fig:09bw}). The errors are reported in brackets}
 \begin{center}\tabcolsep=01.mm
  \begin{tabular}{cccccc}
  \hline
  \hline
   ID & U & B & V & R & I \\
 \hline
 1 & 14.48 (.03) & 14.41 (.02) & 13.73 (.01) & 13.36 (.01) & 12.97 (.01) \\
 2 & 16.60 (.03) &  15.64 (.02) & 14.47 (.02) & 13.83 (.01) & 13.23 (.01) \\
 3 & 16.21 (.02) & 15.97 (.02) & 15.15 (.01) & 14.64 (.01) & 14.15 (.01) \\
 4 & 17.01 (.02) & 16.42 (.02) & 15.41 (.01) & 14.85 (.01) & 14.27 (.01) \\
 5  & 17.92 (.02) & 17.36 (.02) & 16.34 (.01) & 15.79 (.02) & 15.21 (.02)  \\
 6 &  17.48 (.02) & 17.02 (.02) & 16.12 (.02) & 15.61 (.02) & 15.10 (.03)\\
 7 &  16.60 (.02) & 16.46  (.02) & 15.68 (.01) & 15.24 (.01) & 14.78 (.01) \\
 8 & 16.23 (.02) & 16.22 (.02) & 15.54 (.01) & 15.17 (.02) & 14.75 (.01) \\
 9 & 15.13 (.02) & 14.39 (.02)& 13.27 (.02) & 12.70 (.02) & 12.05 (.01)\\
 10 & 18.04 (.01) & 17.73 (.02) & 16.86 (.02) & 16.45 (.01)& 16.01 (.02)\\
 11 &  18.90: & 18.86 (.03) & 18.07 (.02) & 17.64 (.02) & 17.24 (.02) \\
 12 &   & 19.85 (.02) & 18.40 (.02) & 17.39 (.01) & 16.57 (.01) \\
 13 &  19.27: & 19.22 (.02) & 18.50 (.03) & 17.99 (.02) & 17.59 (.02) \\
 14 &   & 19.98 (.02) & 19.07 (.03) & 18.62 (.02) & 18.13 (.03) \\
 15 &   18.55: & 18.76 (.02) & 17.94 (.02) & 17.55 (.01) & 17.09 (.02) \\
 16 &   18.80: & 18.65 (.01) & 17.59 (.02) & 17.00 (.01) & 16.45 (.01) \\
 17 &   20.46: & 19.51 (.03) & 18.47 (.02) & 17.82 (.02) & 17.26 (.02) \\
 18 &   & 20.94 (.02) & 19.36 (.03) & 18.34 (.02) & 17.52 (.02) \\
 \hline
\end{tabular}
\end{center}
\label{table:ls}
\end{table}

Photometric zero points and colour terms were computed for all nights through observations of Landolt standard fields \citep{landolt}.
Thirteen of the forty-nine nights were photometric. Using the photometry of these nights we calibrated the magnitudes of a local stellar sequence shown in Fig.~\ref{fig:09bw}. 
Magnitudes of the local sequence stars are reported in Tab.~\ref{table:ls} along with their root mean square (r.m.s., in brackets).
Finally, the average magnitudes of the local sequence stars were used to calibrate the photometric zero points obtained in non-photometric nights. 
The sequence extends from bright (V=13.3) to faint (V=19.4) stars to allow
a good calibration of the SN magnitudes from the luminous peak to the nebular phase.
The brightest stars of the sequence were saturated on long-exposure frames with larger telescopes. Their calibration is based, therefore, on a fewer photometric nights.
The calibrated optical magnitudes of the SN are reported in Tab.~\ref{table:snm}. The discovery magnitude reported in \citet{c1} has been revised and reported in Tab.~\ref{table:snm}. 
There is no evidence of the SN presence on images obtained on August 24, 2010; the values in Tab.~\ref{table:snm} are upper limits computed with artificial stars placed close to SN position. These values are not shown in Fig.~\ref{fig:sn_lc} because they do not impose tight limits to the late phase decline.

Because of the complex background SN magnitudes have been evaluated in two ways: a) through the point spread function (PSF) fitting technique; 
b) with the template subtraction and subsequent PSF fitting technique.
The magnitudes calculated with both methods during the plateau period result in good agreement. 
However, the results differ up to 0.5 mag during the radioactive tail phase
(especially in bluer bands), and preference has been given to the reduction via the template subtraction. 
The values reported in Tab.~\ref{table:snm} after June 26th have been obtained by the latter technique.
The uncertainties reported in Tab.~\ref{table:snm} were estimated by combining in quadrature the errors in the photometric calibration and the error in the PSF fitting through artificial stars.

\begin{table*}\scriptsize
\caption{UBVRI magnitudes of SN 2009bw and assigned errors in brackets.}
\begin{center}
\begin{tabular}{cccccccc}
\hline
\hline
Date & JD & U & B & V & R & I & Inst.\\
yy/mm/dd & (+2400000) & & & & & &\\
\hline
09/03/27 &  54918.42  &    &  &    &  15.88   (.02) &  &  7\\
09/03/30  & 54921.40  &   & 15.49  (.02) & 15.22  (.01)  &  15.09   (.01) & &  6\\
09/03/31  & 54922.24  &   & 15.38  (.02) & 15.01  (.01)  &  14.90   (.01) &  &  6 \\
09/04/01  & 54923.32  &    & 15.18  (.01) & 14.96   (.01)  &  14.72   (.01) &   &  7\\
09/04/03  & 54925.31  &   & &    &  14.55   (.02) & &  7\\
09/04/03  & 54925.50   &  & 15.32  (.10) & 14.88  (.06)  &  14.68  (.01) & 14.44  (.02) &  6\\
09/04/04  & 54926.24  &  14.51   (.05)  & 15.30   (.05)  & 14.95   (.02)   &  14.65   (.01)   &14.37   (.03)  &  9\\
09/04/07  & 54929.33  &  & & 14.99   (.02)   &  14.64    (.02)   &14.33   (.01)  &  8\\
09/04/08  & 54930.46  &  &   &   &  14.65   (.07)  & &  7 \\
09/04/09  & 54931.30  &  14.79   (.07)  & 15.36   (.03)  & 15.03   (.02)  &   14.67  (.02)  & 14.34  (.03)   & 9\\
09/04/12  & 54934.25  &  14.95   (.15)  & 15.57  (.05) &  15.08 (.05)   &  14.72   (.03)  & 14.40   (.03)  &  9 \\
09/04/12  & 54934.31 &   15.11  (.03) & 15.55  (.02) & 15.18  (.02)  &  14.74   (.02) & 14.49  (.02)  & 1\\
09/04/14  & 54936.40   & 15.49  (.02) & 15.73  (.02) & 15.20  (.02)  &  14.77   (.02) & 14.50  (.02) &  2 \\
09/04/14  & 54936.45   &  & &    &  14.77  (.18) &&  7\\
09/04/15  & 54937.39  &  15.52  (.02) & 15.82  (.02)  &15.17  (.01)  &  14.78   (.02) & 14.50  (.02) &  4 \\
09/04/19  & 54941.26  &  15.85   (.11) &  15.96  (.03)  & 15.25   (.02)   &  14.81    (.02) &  14.45   (.03)   & 9\\
09/04/19  & 54941.39  &  16.05  (.02) & 15.98  (.02) & 15.28  (.02)  &  14.80   (.02) & 14.48  (.02) &  4\\ 
09/04/22  & 54944.28  &  16.29   (.13)  & 16.16   (.05)  & 15.30   (.03)   &  14.88    (.02) &  14.49   (.03)  &  9\\
09/04/22  & 54944.34  &   &   &    &  14.81 (.01) & &   7\\
09/04/23  & 54945.28  &  16.49  (.03) & 16.34  (.05) & 15.29  (.02)  &  14.86   (.02) & 14.51  (.02) &  6\\  
09/04/26  & 54948.28  &    & 16.36   (.04)  & 15.40   (.03)   &  14.94    (.02)  & 14.53   (.03)  &  9\\
09/04/28  & 54950.34  &  16.54  (.19) & 16.26  (.10) & 15.40  (.20)  &  14.93  (.12) & 14.61  (.12) &  1\\ 
09/04/29  & 54951.29  &   & 16.44   (.04)  & 15.47   (.03)   &  15.02    (.02) &  14.58   (.02)  &  9\\
09/04/30  & 54952.38  &  &     &   &  14.97   (.25) &  &  7\\
09/05/03  & 54955.33  &  17.27  (.03) & 16.70  (.02) & 15.57  (.02)  &  15.03   (.02) & 14.65  (.02)  & 1\\
09/05/03  & 54955.38   & 17.26  (.04) & 16.75  (.02) & 15.60  (.02)   & 15.10   (.04) & 14.69  (.02) &  3  \\    
09/05/04  & 54956.28  &    & 16.81   (.06)  & 15.61   (.04)   &  15.03   (.03)  & 14.57  (.03)  &  9\\
09/05/08  & 54960.44  &   &  &     &  15.01   (.13) &  &  7\\
09/05/14  & 54966.45  &    &  &    &  15.11    (.05) &   &  7   \\         
09/05/18  & 54970.34  &   & 17.08   (.07)  & 15.79   (.03)  &   15.16    (.03) &  14.65   (.03)  &  9\\
09/05/18  & 54970.44  &    &   &    &  15.20    (.15) &   &  7\\
09/05/21  & 54973.42  &  18.19  (.15) & 17.01  (.05) &15.76  (.02)  &  15.19   (.02) & 14.70  (.03) &  5\\
09/05/25  & 54977.39  &  18.26  (.19) & 17.27  (.04) & 15.83  (.02)   & 15.17   (.02) & 14.74  (.02) &  5\\
09/05/25  & 54977.42  &    & &    &  15.17   (.30) &  &   7\\
09/05/26  & 54978.33  &    &   & 15.78   (.15) &    15.17    (.10)&   14.66 (.15) &  9\\
09/05/30 &  54982.63   & 18.28  (.09) & 17.29  (.02) & 15.75  (.02)  &  15.17   (.02) & 14.73  (.02) &  1  \\
09/05/31	&   54983.32  & &   &     &  15.22    (.15) &  14.70 (.16) & 9\\
09/06/26  & 55009.58  &   & 17.63  (.02) & 15.94  (.02)  &  15.31   (.02) & 14.82  (.02)  & 1\\
09/07/04	&  55017.46  &   &  &  16.04 (.15)   &    15.31 (.12)  &    &   10\\
09/07/19 &  55032.68  &   & 18.15  (.04) & 16.29  (.02)  &  15.60   (.02) & 15.12  (.02) &  3\\
09/07/23  & 55036.58   &  & 18.26  (.08) & 16.38  (.02)  &  15.62   (.02) & 15.13  (.02) &  5\\
09/07/26  & 55039.43  &    & &  &  15.71   (.05) &  &  7\\
09/07/30 &   55043.42  &  & 18.32 (.17)  &   16.57 (.10)    &   15.76 (.10)    &  15.30 (.10)  & 9\\
09/07/30 &  55043.65  &  & 18.43  (.03)  &16.52  (.03) &   15.74   (.03) & 15.23  (.03)  & 1 \\   
09/08/05  &    55049.35  &   &18.62 (.30)  &   16.93 (.10)     &  16.11 (.10)   &   15.61 (.10)   &  9\\
09/08/10  &   55054.40  &   &   &   18.52 (.30)     &  17.54 (.16)    &  17.00 (.15)    &  9\\
09/08/12  &   55056.43  &  &   &  18.78 (.20)  &     17.90 (.18)    &  17.35 (.30) &  9\\
09/08/12  & 55056.54   & 20.04  (.40) & 20.79  (.09) & 18.90  (.10)  &  17.93   (.06) & 17.36  (.08)  & 1\\	 
09/08/13 &  55057.59  &  21.18  (.38) & 20.74  (.09) & 18.96  (.12)  &  18.01   (.04) & 17.41 (.07)   & 2 \\
09/08/20  & 55063.61  &  & 20.80  (.21) & 19.23  (.11) &  18.12   (.17) & 17.47  (.13) &  5\\ 
09/08/23  &   55067.29  &   &  &    &     18.03 (.15)   &     17.59 (.30) &  9\\
09/08/29  &  55073.55  &  &  &  19.06 (.15)     &   18.05 (.10) &	&  11\\
09/08/30  &  55074.52  &  &   &     &   18.05 (.15)  &	&  11\\
09/08/30  & 55074.57  &  & 20.81  (.04) & 19.14  (.13)  &  18.19   (.08) & 17.57  (.08)  & 1  \\ 
09/09/04  & 55079.44  &  & 21.02  (.05) & 19.19  (.04)  &  18.14   (.04) & 17.55  (.07) &  1 \\
09/09/04  &  55079.53  &   &    &  19.15 (.16)    &   18.14 (.12)    &  17.51 (.27)   &  11\\
09/09/05  &   55080.47  &  &  &     &    18.12 (.12)   &     &   11\\
09/09/06  &  55081.43   & &  &       &    18.15 (.12)   &   &   11\\
09/10/12 &  55116.74  &   & 21.03  (.11) & 19.48  (.29)  &  18.51   (.57) & 18.10  (.15) &  3\\
09/10/18  & 55123.62   &  &  &19.62  (.13)   & 18.70   (.11) & 18.20  (.12) &  5\\ 
09/11/07 &   55143.50  &   &   &   20.09 (.27)    &   19.01 (.15)	&    	  &    11\\
09/11/09 &   55145.47   &  &  &  20.13 (.21)   &    19.05 (.18)	  &	 &     11\\
09/11/10 &  55146.46   &  & 21.41  (.12) & 20.11  (.19)  &  19.05   (.56) & 18.68  (.22) &  3\\
09/11/17  &   55153.42  &   &  &   20.36 (.21)    &   19.22 (.12)  &   &   11\\
09/11/22  & 55157.85  &  &  & 20.40  (.21)   & 19.36   (.24) & 18.78  (.25) &  1\\
10/08/24  & 55433.69 & & & $>$ 21.9 & $>$ 20.3 & $>$ 19.5 & 3\\
\hline
\end{tabular}
\end{center}
1 = CAHA,  2 = NOT, 3 = TNG, 4 = LT, 5 = Copernico, 6 = SAO-RAS, 7 = THO, 8 = S50, 9 = M70, 10 = C50, 11 = C60. The telescopes abbreviations are the same of Tab.~\ref{table:cht}. 
\label{table:snm}
\end{table*}%

Near infrared (NIR) photometry was obtained at two epochs with NICS  (cfr. Tab.~\ref{table:cht}) mounted at the 3.5m Telescopio Nazionale Galileo (TNG).
The NIR images of the SN field were obtained combining several, sky-subtracted, dithered exposures. Photometric calibration was achieved relative to 2MASS photometry of the same local sequence stars as used for the optical calibration. The NIR magnitudes of \bw\/ are listed in Tab.~\ref{table:snir}.

Ultraviolet \citep[uvw2,uvm2,uvw1; see][]{swift} observations, obtained by UVOT on board of the SWIFT satellite are available for four epochs in a period of 9d. 
We reduced these data using the HEASARC\footnote{NASA's High Energy Astrophysics Science Archive Research Center} software. For each epoch all images were co-added, and then reduced following the guidelines presented by \citet{swift}. Aperture magnitudes are reported on Tab.~\ref{table:swf}.
 
In Sect.~\ref{sec:dis} we will discuss also the implications of an 8 ksec exposure obtained with Swift-XRT which has provided an upper limit of 7.9$\times10^{-14}$erg cm$^{-2}$ s$^{-1}$ (over the energy range 0.3-10 KeV) in an aperture of 9 XRT pixels ($\sim$21") centered at the position of \bw.

\subsection{Photometric evolution}\label{sec:pe}
In Fig.~\ref{fig:sn_lc} the \textit{uvw2,uvm2,uvw1,U,B,V,R,I,J,H,K$^{\prime}$} light curves of \bw\/ are plotted.
The B-, V-, I- and more clearly R-band light curves show a slow rise to the peak, estimated to be occurred around the JD 2454923.5$\pm$1.0 in the B band. It is likely that thus the discovery of the SN happened close to the shock breakout.
The early discovery is also consistent with the phases
derived for the first spectra of \bw\/ as deduced from the comparison of the early spectra with a library of supernova spectra performed with the ''GELATO'' code \citep{avik}. Therefore,
hereafter we will adopt JD 2454916.5$\pm$3 (March 25.0 UT) as an estimate for the epoch of shock breakout.

An initial slow decline of $\sim$30d, during which the SN decreases by about 0.5 mag in the V and R bands, is followed by a plateau lasting for 50 d to 100 d. The plateau is flat from V to I ($m_{R} \sim 15.0$, i.e. $M_{R} \sim -17.4$, cfr. Sect.~\ref{sec:red}), possibly shorter and inclined in blue bands. 
The plateau luminosity of \bw\/ is
more luminous than that of common SNe IIP  \citep[more than 1 mag, see][]{p1,richardson} and similar to those of SNe 1992H \citep{92h}, 2004et \citep{04et} and 2007od \citep{07od}. 

As shown in Fig.~\ref{fig:sn_lc}, after $\sim$110d {\it BVRI} light curves show a decline from the plateau. 
Though very fast ($\sim$2.2 mag in only 13d) the jump is less pronounced than in other SNe IIP e.g. 6 mag in SN 2007od \citep{07od}, $>$3.5 mag in SN 1994W \citep{94w}, 3.6 mag in the low luminosity SN 2005cs \citep{05cs2}.


After the drop, the light curve settles onto the radioactive tail. The late time decline rates in various bands are reported in Tab.~\ref{table:main}. Although with some variations, these share an overall similarity with those of most SNe IIP \citep[e.g.][]{t1,p1} and are 
close to 0.98 \mcento\/ , the decay rate of \co\/ to \fe\/ (corresponding to a lifetime of 111.26d).

\begin{figure}
\includegraphics[width=\columnwidth]{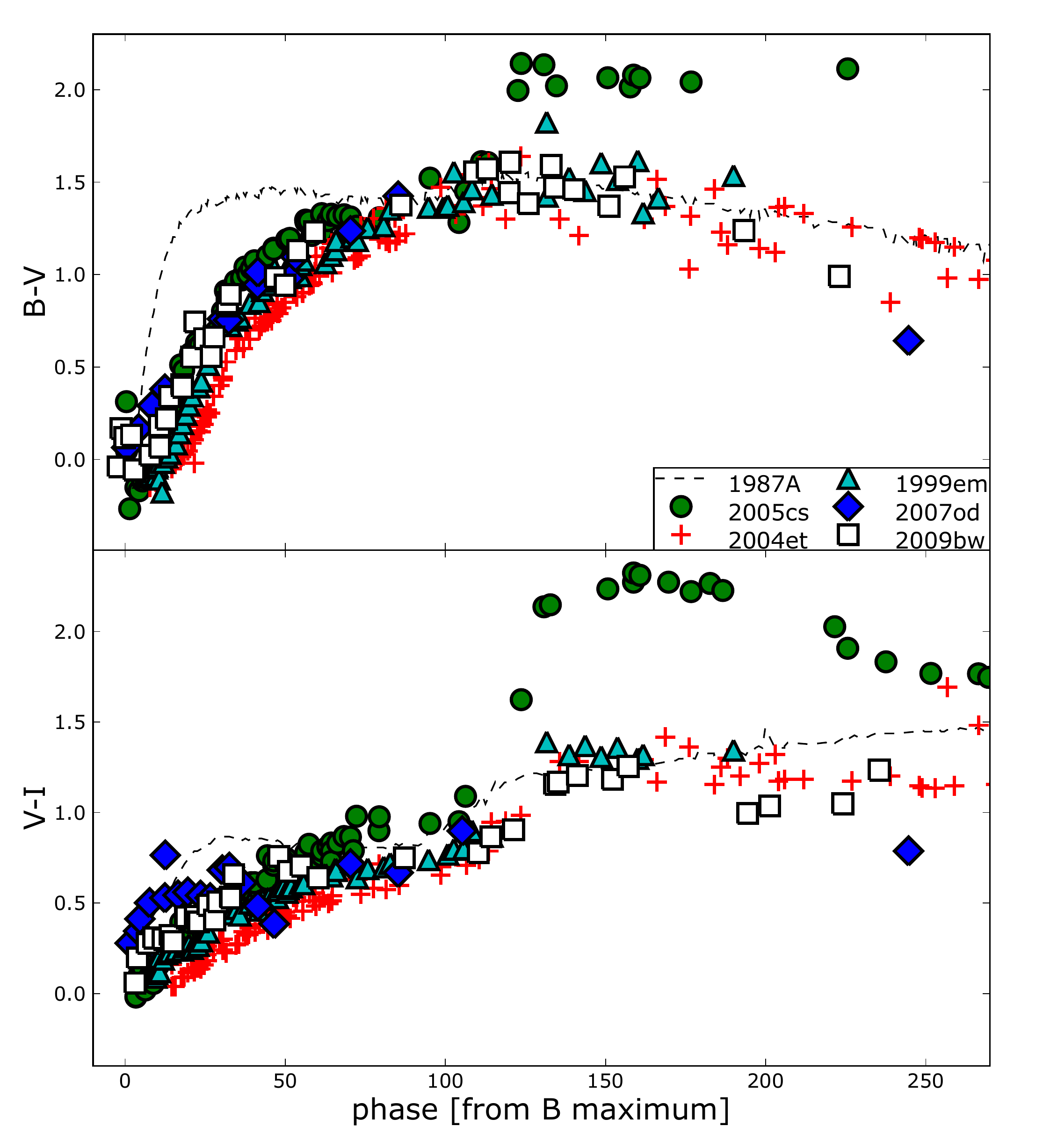}
\caption{Comparison of the dereddened colours of SN 2009bw and those of  SNe 1987A, 2005cs, 1999em, 2004et and 2007od. The phase of SN 1987A is with respect to the explosion date.}
\label{fig:col}
\end{figure}

In Fig.~\ref{fig:col} we show the time evolution of the B~--~V and V~--~I colour curves of \bw\/  together with those of SNe 1987A, 2005cs, 2004et, 1999em and 2007od, dereddened according to the values of Tab.~\ref{table:snc}.

All typical SNe IIP show a similar colour evolution with a rapid increase of B--V as the SN envelope expands and cools. After about 40d the colour changes more slowly as the rate of cooling decreases.  \bw\/ follows the general behavior. The only exception to this smooth trend is SN 2005cs which shows a red jump about at 120d, that seems a characteristic of low-luminosity SNe II \citep{pa1}.
The B--V trend of \bw\/  during the early days seems redder than that of normal SNe IIP such as SN~1999em, and closer to that of SN~2007od.
The last two points suggest a 
blue excess, as does the SN 2007od, but the errors of these points are large and we will not consider it any further.
The V--I colour evolution of \bw\/ is similar to that of other type IIP SNe (cfr.  Fig.~\ref{fig:col}).

\begin{table}
\caption{JHK$^{\prime}$ TNG magnitudes of SN 2009bw and assigned errors.}
\begin{center}\tabcolsep=1.3mm
\begin{tabular}{cccccc}
\hline
\hline
Date & JD & J & H & K$^{\prime}$ \\
yy/mm/dd & (+2400000) & & &\\
\hline
09/10/07 & 55112.49 & 18.17 (.18)  &  17.99 (.15)	&   17.59   (.13)\\
09/12/05 & 55171.50 &    & 	&   17.89   (.08) \\
 \hline
  \end{tabular}
\end{center}
\label{table:snir}
\end{table}%

\begin{table}
\caption{Swift magnitudes of SN 2009bw and assigned errors.}
\begin{center}\tabcolsep=1mm
\begin{tabular}{ccccc}
\hline
\hline
Date & JD & uvw2 & uvm2 & uvw1 \\
yy/mm/dd & (+2400000) & & & \\
\hline
09/04/01 & 54922.51  &   14.93 (.06) & 14.72 (.07) & 14.50 (.10)\\
09/04/02 & 54923.51  &   15.16 (.06) & 14.92 (.08) &  \\
09/04/03 & 54924.52   &   & 15.13 (.09) &  \\ 
09/04/09 & 54931.05  &   15.39 (.06) &&  \\ 
\hline
\end{tabular}
\end{center}
\label{table:swf}
\end{table}%

\begin{table}
\caption{Main data of \bw\/.}
\begin{center}\tabcolsep=0.95mm
\begin{tabular}{lcc}
\hline
position (2000.0)	&03$^h$56$^m$02$^s$.92	&+72$^o$55$^m$40$^s$.9 \\
parent galaxy morphology			& \multicolumn{2}{c}{UGC~2890,  Sdm pec:}\\
offset wrt  nucleus 		& 11$\arcsec$E		& 22$\arcsec$N 	\\
adopted distance modulus	&  $\mu=31.53\pm0.15$	&	\\
SN heliocentric velocity	&  $1155\pm6$ \kms	&	\\
adopted reddening		& E$_{g}$(B-V)$=0.23$	& E$_{tot}$(B-V)$=0.31$\\
\hline
\end{tabular}
\begin{tabular}{lccc}
	& peak time		& peak observed	& peak absolute  \\
	&  (JD 2454000+)	&  magnitude		&  magnitude	\\
B	& $923\pm1$		&$15.18\pm0.01$		&$-17.70\pm 0.15$\\
V	& $925\pm1$		&$14.88\pm0.06$		&$-17.67\pm 0.16$\\
R	& $925 \pm1$		&$14.55\pm0.02$ 	&$-17.82\pm 0.15$ \\
I	& $929\pm1$		&$14.33\pm0.01$	&$-17.83\pm 0.15$\\
UBVRI& $925\pm2$	& \multicolumn{2}{c}{L$_{bol}=2.6 \times 10^{42}$ erg s$^{-1}$}  \\
	&				&				&				\\
rise to R max & $\sim8$ days	& 			&			\\
explosion day & $\sim916.5\pm3$ & \multicolumn{2}{c}{$\sim25$ Mar. 2009}			\\
\hline
\end{tabular}
\begin{tabular}{p{4.4cm}cc}
	& late time decline	&interval  \\
	& [mag(100d)$^{-1}$]		& [days]  \\
V	& 	1.00			& 139--239\\
R	& 	1.09			& 139--239\\
I	& 	1.13			& 139--239\\
UBVRI(tot)&    1.06		& 139--239\\
UBVRI(1seg)&    0.70	& 139--156  \\
UBVRI(2seg)&    1.16 	& 161--239  \\
\hline
\end{tabular}
\begin{tabular}{p{6cm}c}
M($^{56}$Ni)			&  0.022 M$_{\odot}$\\
M(ejecta)			& 8.3 -- 12  M$_{\odot}$ \\
explosion energy	&  0.3 $\times 10^{51}$  ergs\\
\hline
\end{tabular}
\end{center}
\label{table:main}
\end{table}

\begin{table}
  \caption{Main parameters of SNe II used in the comparisons with \bw.}
  \begin{center}
  \begin{tabular}{ccccc}
  \hline
  \hline
  SN & $\mu^{*}$ & \ebv & Parent Galaxy & References  \\
 \hline
 1987A & 18.49 & 0.195 & LMC & 1 \\
 1992H & 32.38  & 0.027 & NGC 5377& 2 \\
 1998S & 31.08 & 0.232 & NGC 3877&3 \\
 1999em & 29.47 & 0.1 &NGC 1637& 4,5 \\
 2004et & 28.85 & 0.41 & NGC 6946 & 6 \\
 2005cs & 29.62 & 0.05 & M 51& 7 \\
 2006bp & 31.44 & 0.031 &NGC 3953 & 8\\
 2007od & 32.05 & 0.038 & UGC 12846 & 9\\
  \hline
\end{tabular}
\end{center}
$^{*}$ In the H$_{0}= 73$ \kms\/ Mpc$^{-1}$ distance scale\\
REFERENCES: 1 - \citet{87a}, 2 - \citet{92h}, 
3 - \citet{fa01}, 4 - \citet{99em}, 5 - \citet{99em2}, 6 - \citet{04et},  7 - \citet{05cs2}, 8 - \citet{qu07}, 9 - \citet{07od}
\label{table:snc}
\end{table}
  
\subsection{Reddening and absolute magnitude}\label{sec:red}
The Galactic reddening to UGC 2890 is E$_{g}$(B-V) = 0.231 mag (A$_{g}(B)$ = 0.996 mag) according to \citet{ext}, a relative large value consistent with the low Galactic latitude of UGC 2890 (b$_2$=14$^{o}$.7, LEDA). In the optical spectra of \bw\/ the absorption features due to interstellar NaID ($\lambda\lambda$5890,5896) of the Galaxy are present, with an average EW$_{g}$(NaID)$\sim$1.37 \AA\/, as determined by our best resolution spectra (cfr. Sect.~\ref{sec:spec}). According to \citet{t2} this corresponds to a Galactic reddening E$_{g}$(B-V)$\sim$0.22 mag (A$_{g}$(B)$\sim$0.92 mag). This is in good agreement with \citet{ext}.
With the same method we can estimate the reddening inside the parent galaxy.
Interstellar NaID components within the host galaxy have been measured with an average EW$_{i}$(NaID)$\sim$ 0.52 \AA\/ that corresponds to a low internal reddening E$_{i}$(B-V)$\sim$0.08 mag or A$_{i}$(B)$\sim$0.35 mag.
We should warn the reader that recently \citet{poz11} have suggested that NaID lines in SNIa spectra are poor tracers of dust extinction.
However, having no alternative throughout this paper we have adopted a total reddening E$_{tot}$(B-V) = 0.31$\pm0.03$ mag.

NED provides a heliocentric radial velocity of $v_{hel}(UGC2890) = 1155 \pm 6$ \kms. Adopting H$_{0} = 73$ \kms\/ Mpc$^{-1}$ and a velocity corrected for the Virgo infall of $v_{Virgo}=1473\pm16$ \kms\/  \citep{mould} we obtain a distance modulus $\mu$ = 31.53 $\pm$ 0.15 mag which will be used throughout this paper. 
\citet{tully} provided a $\mu$ = 30.23 $\pm$ 0.40 mag (H$_{0}$ = 72 \kms\/ Mpc$^{-1}$), this  would make \bw\/ more than 1 mag fainter, however it is inconsistent with the identified galaxy lines.

Assuming the above distance and extinction values, we find $M^{max}_{B} =-17.72\pm0.15$, $M^{max}_{V} =-17.67\pm0.16$, $M^{max}_{R} =-17.82\pm0.15$ and $M^{max}_{I} =-17.83\pm0.15$, where the reported errors include the uncertainties on the adopted distance modulus, reddening and magnitude measurements.

\subsection{Bolometric light curve and \ni\/ mass}\label{sec:bol}

Unfortunately, because of the lack of simultaneous optical--NIR observations it is impossible to obtain a true bolometric light curve for \bw. In fact extended coverage from B to K$^{\prime}$ is available only in the late radioactive tail. 
In order to obtain the bolometric light curve, we convert only {\it UBVRI} broadband magnitudes (corrected for the adopted extinction, cfr. Sect.~\ref{sec:red}) into fluxes at the effective wavelengths and integrate them over the entire range (flux integration limits equal to zero).
Flux was then converted to luminosity using the distance adopted in Sect.~\ref{sec:red}. 
The emitted flux was computed at phases in which R observations were available.
When simultaneous observations in a bandpass were unavailable, the magnitudes were interpolated from the light curves using low-order polynomials, or extrapolated. 
The peak of the bolometric light curve is reached close to the R-band maximum on JD$^{bol}_{max}$ = 2454925.3$\pm$ 2.0 at a luminosity L$_{bol}$ = 2.6 x 10$^{42}$ erg s$^{-1}$. 
\begin{figure}
\includegraphics[width=\columnwidth]{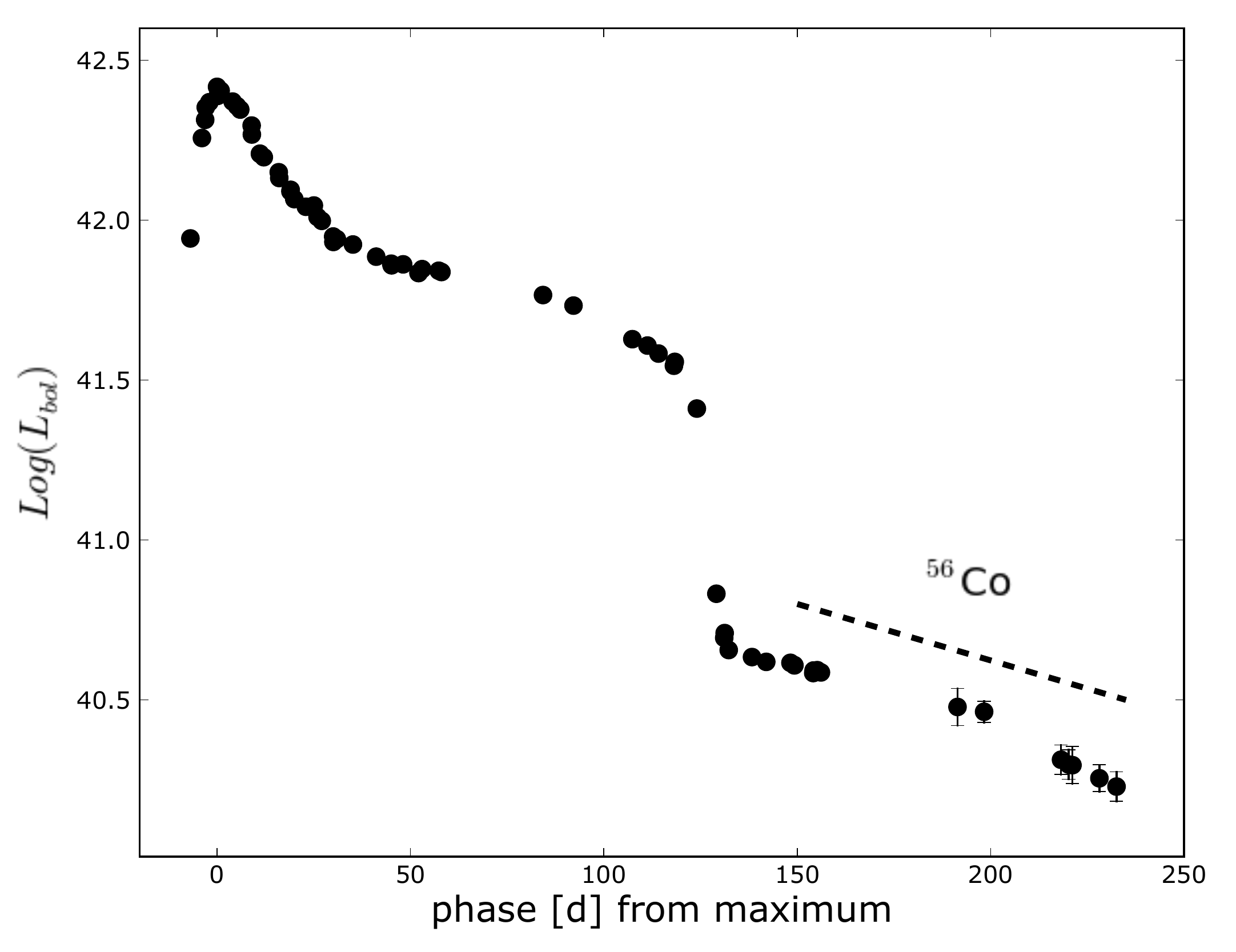}
\caption{{\it UBVRI} bolometric light curve of \bw\/. The slope of $^{56}$Co to $^{56}$Fe decay is also displayed for comparison. Distance modulus and reddening are reported in Tab.~\ref{table:main}.} 
\label{fig:09_bol}
\end{figure}


The nebular tail of the bolometric light curve declines at a rate $\gamma \sim 1.06$ mag $(100d)^{-1}$, measured from 138d--239d since explosion, which matches the decay of $^{56}$Co to $^{56}$Fe suggesting complete $\gamma$-ray trapping. 
However, we noticed that the tail can be divided into two different segments: a flatter one from 138d to 156d with $\gamma \sim 0.70$ mag $(100d)^{-1}$ resembling that of the early tail of \em\/ and present in many SNe, and a second one (161d--239d) 
with $\gamma \sim 1.16$ mag $(100d)^{-1}$. 
The last two points at $\sim$234-240d seem to suggest a further increase of the decline rate, maybe due to dust formation. However, due to the relatively large errors this is not significant and we will not elaborate it any further. 

\begin{figure*}
\includegraphics[width=18cm]{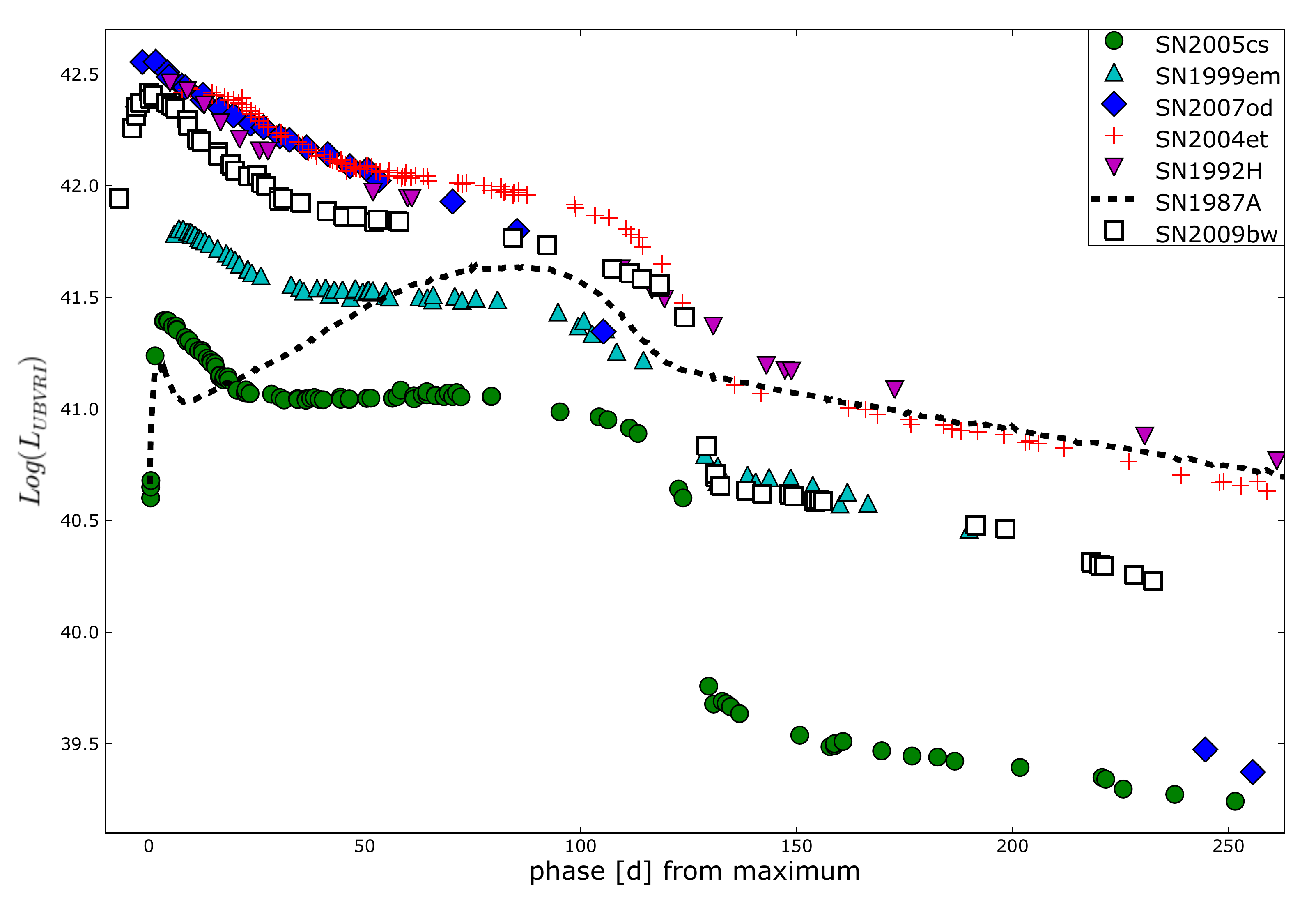}
\caption{Comparison of bolometric ({\it UBVRI}) light curves of SN 2009bw with those of other Type II SNe. The distances and reddenings for the various objects are reported on Tab.~\ref{table:snc}. 
Small misalignments in the epoch of the maximum are due to different shifts between the maxima of the band adopted as reference and the bolometric one.} 
\label{fig:cfr_bol}
\end{figure*}

In Fig.~\ref{fig:cfr_bol} we compare the bolometric ({\it UBVRI}) light curve of \bw\/ with those of other SNe
reported in Tab.~\ref{table:snc}.
The early luminosity of \bw\/ is slightly smaller than those of the luminous SNe 2007od, 2004et, 1992H. The duration of the plateau resembles those of SN 1992H and SN 2004et, whilst it is longer than that of the peculiar SN 2007od. The peak seems slightly broader than those of other SNe IIP including 
SN 2007od. 
The luminosity drop from the plateau to the tail of \bw\/ appears intermediate between the high value for 2007od and the small value for 1999em.

The $^{56}$Ni mass ejected by \bw\/ has been derived by comparing its bolometric light curve to that of SN 1987A assuming similar $\gamma$-ray deposition fraction
\begin{equation}
M(^{56}Ni)_{09bw}= M(^{56}Ni)_{87A}\times\frac{L_{09bw}}{L_{87A}} M_{\odot} 
\end{equation}
where  M($^{56}$Ni)$_{87A}$ = 0.075 $\pm$ 0.005 M$_{\odot}$ is the mass of $^{56}$Ni produced by SN 1987A \citep{87a1}, and L$_{87A}$ is the {\it UBVRI} luminosity of 1987A at a comparable epoch.
The comparison, performed between 161d and 205d from explosion, gives M($^{56}$Ni)$_{09bw}$ $\sim$ 0.022 M$_{\odot}$.
Indeed L$_{bol}(09bw)$ on the tail is similar to that of SN 1999em that ejected $\sim 0.022$ M$_{\odot}$ of $^{56}$Ni \citep{99em}.
We have crosschecked this result by using the formula from \citet{hamuy} and assuming that $\gamma$-rays resulting from the \co\/ decay are fully thermalized at this epoch 
\begin{equation}\label{eq}
\textstyle M(^{56}Ni)_{09bw} = \left(7.866\times10^{-44}\right) L_{t}exp\left[\frac{(t-t_{0})/(1+z)-6.1}{111.26}\right] M_{\odot}
\end{equation}
where t$_{0}$ is the explosion time,  6.1d is the half-life of $^{56}$Ni and 111.26d is the \textit{e}-folding time of the $^{56}$Co decay, that release 1.71 MeV and 3.57 MeV, respectively, in the form of $\gamma$-rays \citep{capdecay,w2}. This method provides M(\ni)$\sim$ 0.021 \M, consistent with the preceding value within the uncertainties.
The single epoch (161d) of NIR observations shows a NIR contribution of $\sim$20\%, which is similar to that of the other IIP SNe \citep[see Fig.7 of][]{07od}.

Good sampling of the end of the plateau phase and the beginning of the radioactive tail allows us to estimate the steepness function {\it S} (maximum value at the transition phase of the first derivative of the plateau absolute visual magnitude) and in turn the $^{56}$Ni mass using the method devised by \citet{elm}.
We evaluated {\it S} = 0.57 corresponding to 0.002 M$_{\odot}$ of $^{56}$Ni, typical of faint CC-SNe and very different from the previous estimates. The {\it S} value is larger than in all SNe of the sample of \citet{elm} and extremely different to those of similar luminosity SNe 1992H ({\it S} = 0.048) and 1999em ({\it S} = 0.118). This result suggests that in this case the anti-correlation between steepness function and $^{56}$Ni mass does not work, implying something uncommon in the high photospheric luminosity or a transition masked by some effect. We will address this issue in Sect.~\ref{sec:dis}.


\section{Spectroscopy}\label{sec:spec}

The spectroscopic monitoring of \bw\/ was carried out with several telescopes over a period of six months. The journal of spectroscopic observations is reported in Tab.~\ref{table:sp}.

\begin{figure*}
\includegraphics[width=18cm]{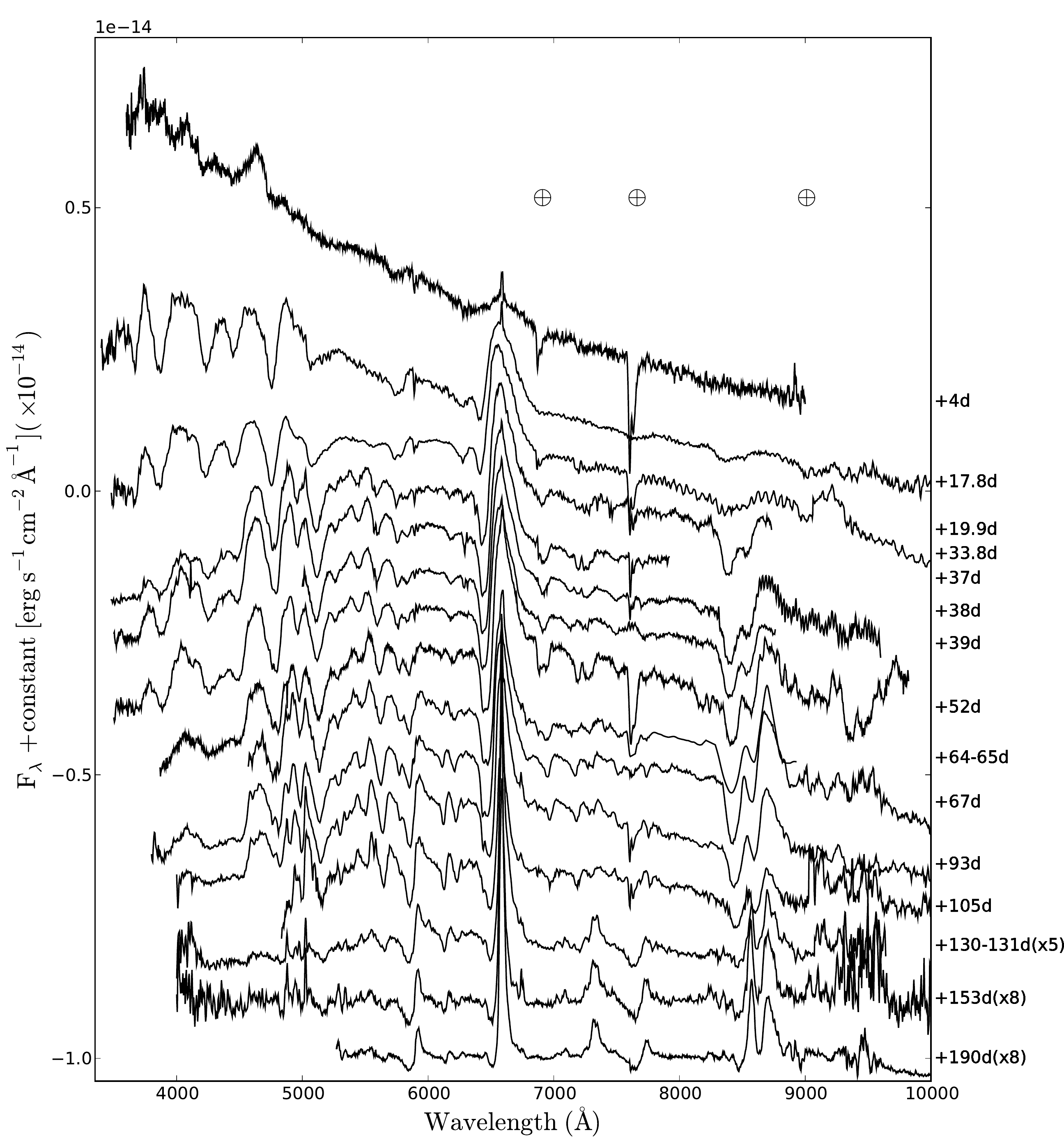}
\caption{The overall spectral evolution of SN 2009bw. Wavelengths are in the observer rest frame. The phase reported for each spectrum is relative to the explosion date (JD 2454916.5). The $\oplus$ symbols mark the positions of the most important telluric absorptions. The ordinate refers to the first spectrum. The second spectrum is shifted downwards by $2\times 10^{-15}$ units, the third by $3.3\times 10^{-15}$ units with respects to the second, others by $1.5\times 10^{-15}$ erg s$^{-1}$ cm$^{-2}$ \AA\/$^{-1}$ with respect to the previous.} 
\label{fig:spec_ev}
\end{figure*}

\subsection{Spectra reduction}\label{sec:sr}

Spectra were reduced (including trimming, overscan, bias correction and flat-fielding) using standard routines of IRAF. An optimal extraction of the spectra was adopted to improve the signal-to-noise rate (S/N). Wavelength calibration was performed through spectra of comparison lamps acquired with the same configurations as the SN observations. Atmospheric extinction correction was based on tabulated extinction coefficients for each telescope site. Flux calibration was done using spectrophotometric standard stars observed with the same set-up as SN in the same nights. Absolute flux calibration was checked by comparison with the photometry, integrating the spectral flux transmitted by standard {\it BVRI} filters and, when necessary, adjusting it by a multiplicative factor. 
The resulting flux calibration is accurate to within approximately 0.1 of magnitude. The spectral resolutions in Tab.~\ref{table:sp} were estimated from the full widths at half maximum (FWHM) of the night sky lines.  We used the spectra of standard stars to remove when possible telluric features in the SN spectra. The regions of the strongest atmosphere features are marked in Fig.~\ref{fig:spec_ev}. Spectra of similar quality obtained in the same night with the same telescope have been combined to increase the S/N. 

The spectrum of  May 28 (64d) revealed some problems in the flux calibration of the blue side. The spectral continuum was forced to follow the spectral energy distribution (SED) derived  fitting with a low order polynomial function the photometry obtained in the same night.
This spectrum has not been used in the estimate of the temperature in Sect.~\ref{sec:ev}.
Since no significant evolution is expected, the spectra at 64d and 65d, as well as those at 130d and 131d have been combined in Fig.~\ref{fig:spec_ev}.
 
For \bw\/ a single NIR spectrum was collected at  a phase at 20d, which was reduced and extracted through standard routines of IRAF. As for the optical, the spectrum was calibrated in wavelength through spectra of comparison lamps acquired with the same configuration of the SN observation. First order flux calibration was obtained using telluric A0 standard stars taken in the same night with the SN set-up. The prominent telluric bands were identified through the spectra of the standard star and thus removed. The combined optical and NIR spectrum on day 20 is shown in Fig.~\ref{fig:spec_nir}.

\begin{table*}
  \caption{Journal of spectroscopic observations of SN 2009bw.}
  \begin{center}
  \begin{tabular}{cccccc}
  \hline
  \hline
  Date & JD & Phase$^{*}$ & Instrumental & Range & Resolution$^{\dagger}$ \\
  & +2400000 & (days) & configuration &(\AA) & (\AA) \\
 \hline
 09/03/29 & 54920.5 & 4.0 & NOT+ALFOSC+gm4 & 3600-9000 & 13\\
 09/04/12 & 54934.3 &17.8 & CAHA+CAFOS+b200,r200 & 3380-10000 & 10  \\
 09/04/14 & 54936.3 & 19.8 & NOT+ALFOSC+gm4 & 3480-10000 & 13\\
 09/04/15 & 54937.3 & 20.8 & TNG+NICS+ij,hk & 8750-24600 & 18,36\\
 09/04/28 & 54950.3 & 33.8 & CAHA+CAFOS+b200 & 3480-8730 & 10 \\
 09/05/01 & 54953.5 & 37.0 & TNG+DOLORES+LRB &3500-7900 & 15 \\
 09/05/02 & 54954.5 & 38.0 & TNG+DOLORES+LRR & 5000-9600 & 15 \\
 09/05/03 & 54955.3 & 39.0 & CAHA+CAFOS+b200 & 3500-8760 & 10 \\
 09/05/16 & 54968.5 & 52.0 & BTA+SCORPIO+G400& 3870-9830 & 13  \\
 09/05/28 & 54980.5 & 64.0 & Copernico +AFOSC+gm4 & 4500-7760 & 25\\
 09/05/29 & 54981.5 & 65.0 & Copernico +AFOSC+gm2 & 5330-9020 & 37 \\
 09/05/31 & 54983.5 & 67.0 & CAHA+CAFOS+g200 & 3780-10000 & 9\\
 09/06/26 & 54509.5 & 93.0 & CAHA+CAFOS+g200 & 4000-10000 & 10\\
 09/07/18 & 55021.5 & 105.0 & CAHA+CAFOS+g200 & 4800-10000 & 14\\
 09/08/12 & 55046.5 & 130.0 & CAHA+CAFOS+g200 & 4000-9650 & 12\\
 09/08/13 & 55047.5 & 131.0 & NOT+ALFOSC+gm4 & 4200-9100 & 18\\
 09/09/04 & 55069.5 & 153.0 & CAHA+CAFOS+g200 & 4000-10000 & 9 \\
 09/10/11 & 55106.5 & 190.0& TNG+DOLORES+LRR & 5260-10000& 16\\
 \hline
\end{tabular}
 
 * with respect to the explosion epoch (JD 2454916.5)\\
 $\dagger$ as measured from the full-width at half maximum (FWHM) of the night sky lines\\
\end{center}
The abbreviations are the same of Tab.~\ref{table:cht}; in addition BTA = The 6m Big Telescope Alt-Azimutal (Mt. Pastukhova, Russia).
\label{table:sp}
\end{table*}

\subsection{Spectra analysis}\label{sec:sa}

\begin{figure}
\includegraphics[width=\columnwidth]{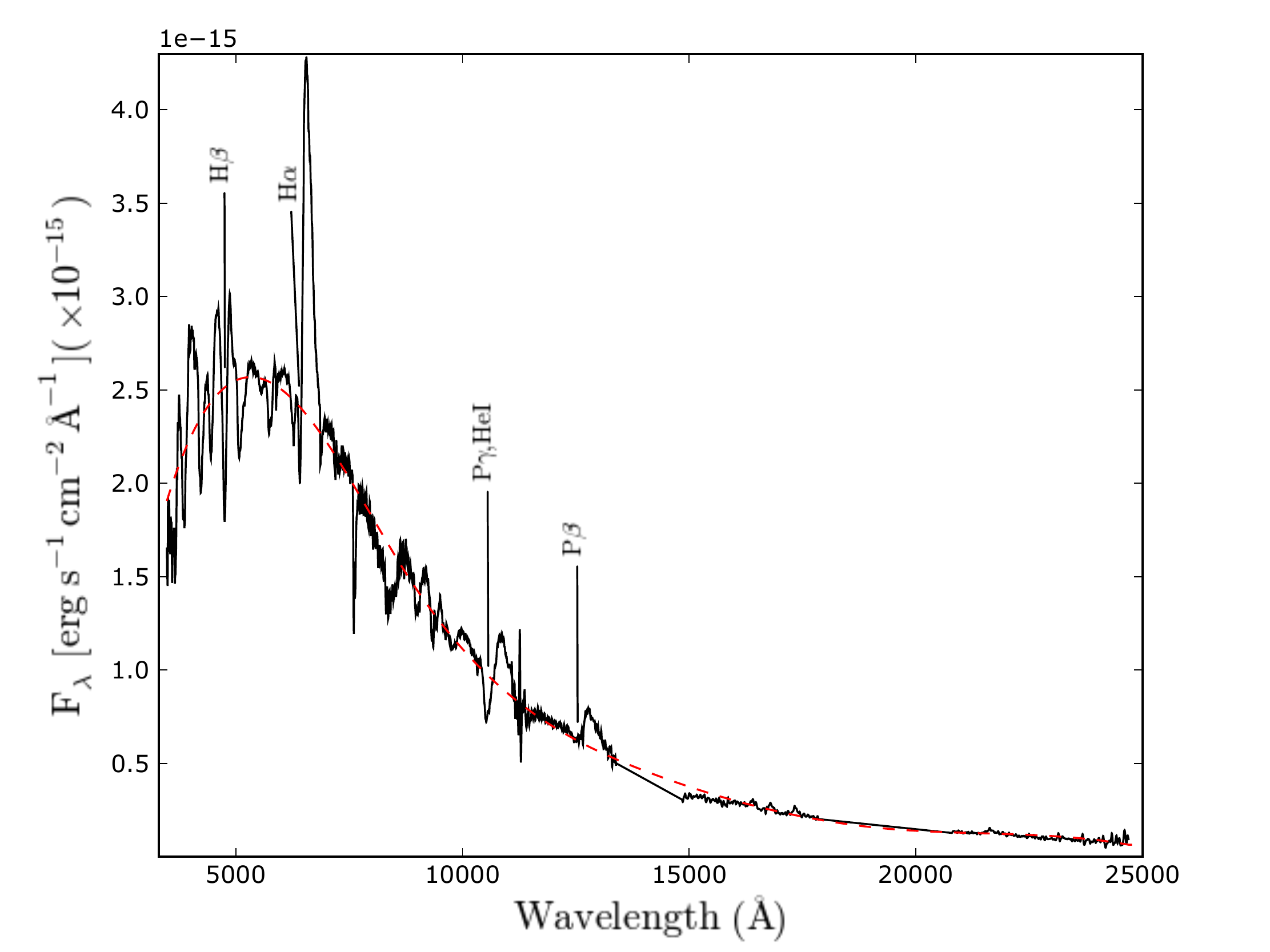}
\caption{Composed spectrum of SN 2009bw, from optical to NIR wavelengths, at $\sim20$d past explosion (JD 2454916.5). Wavelengths are in the observer rest frame. A black body fit at $\sim$7900 K is over-plotted. The regions of the strongest atmospheric features are not shown.} 
\label{fig:spec_nir}
\end{figure}

Fig.~\ref{fig:spec_ev} shows the spectral evolution of \bw\/ from $\sim$2d before maximum to over six months past peak, well sampled both in the photospheric and early nebular stages. 

\begin{figure*}
\includegraphics[width=16cm]{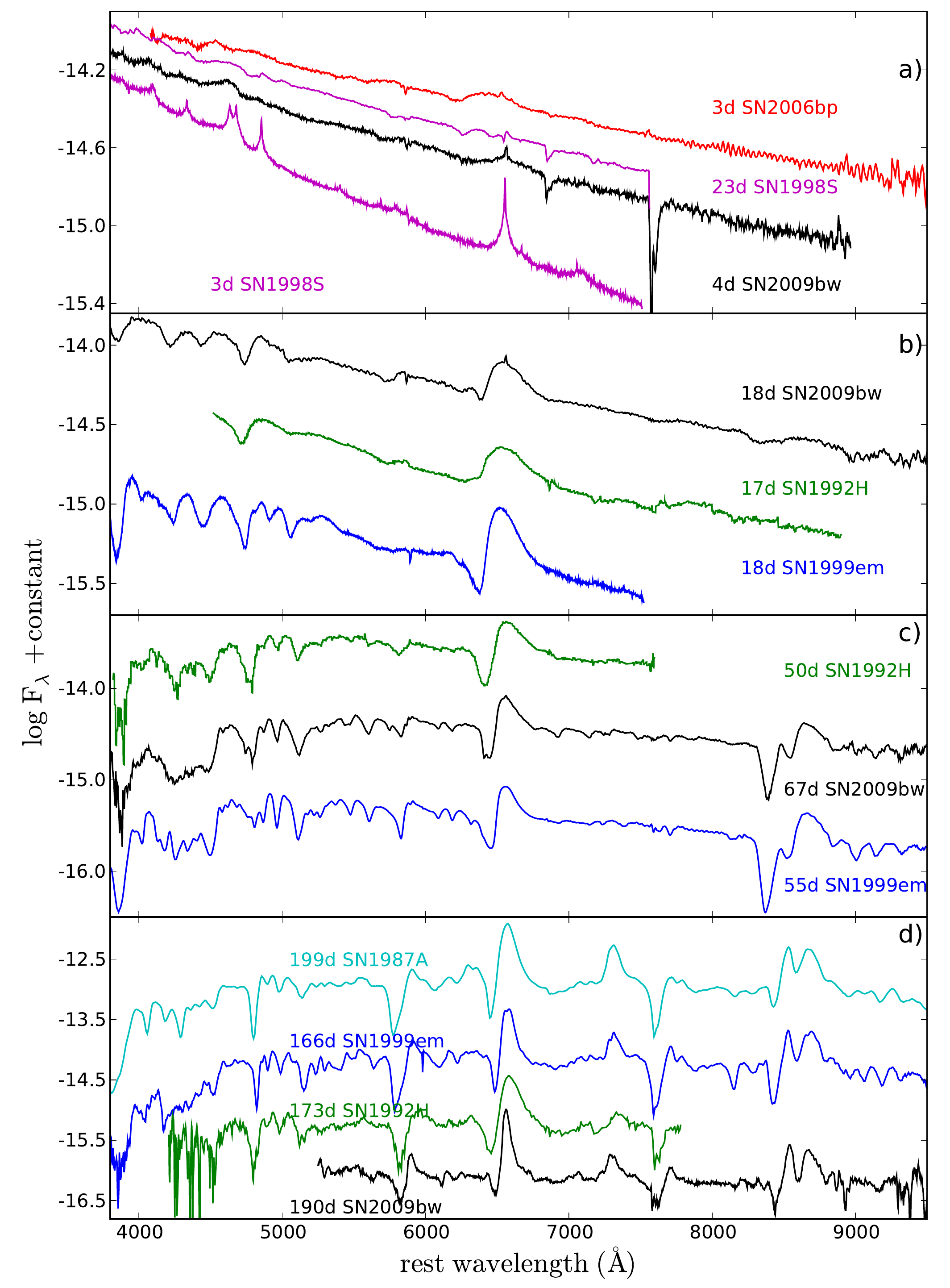}
\caption{Panel a): comparison among spectra of SN 2009bw, SN 1998S and SN2006bp about $\sim$4d past explosion, plus the spectrum of SN 1998S at 23d. Panel b): comparison between spectra of SN 2009bw, SN 1999em and SN 1992H about 18d past explosion. In panel c) spectra of SN 2009bw, SN 1999em and SN 1992H in the plateau phase are compared and in panel d) spectra of SN 2009bw, SN 1987A, SN 1999em and SN 1992H during the nebular phase are shown. For references, see the text and Tab.~\ref{table:snc}, except the SN 1987A spectrum \citep{87a2}.} 
\label{fig:cfr}
\end{figure*}

The first spectrum, used for classification by \citet{c2}, shows a blue continuum typical of SNe II at the same age. An unusual, prominent feature at 4600\AA\/ appears near maximum.
The feature is not present in the subsequent spectrum (18d), and
is most likely related to the SN since it is relatively broad (FWHM$\sim$140\AA\/) and has a short life. No significant emission of the underlying HII region, well seen at other wavelengths, is present in this range even in the 
latest spectra at 190d \citep[N II, N III, O II and forbidden Fe III lines are present in the spectra of typical HII regions at about 4600\AA\/, cfr.][]{HII}.
The FWHM of the 4600\AA\/ feature corresponds to a velocity v$\sim$9100 \kms\/, larger than the velocity of \Ha\/ ($\sim$7000 \kms\/), leading us to explore the possibility of a unusual line formation mechanism or to a very peculiar line blending. 
Similar features seen in ultra-luminous type Ic SNe \citep{pa10,qu11} have been identified as O II, but at wavelengths slightly bluer than this ($\lambda$4414). 
Concerning the opacity, its component related to the bound free process has a saw-tooth behaviour. If we consider a complex atmosphere, formed by H+He+heavy elements (e.g. Fe, Si) with temperature of the order of T$\sim$10$^4$, 
the opacity in a window of the order of 10$^{2}$ \AA\/ around this wavelength is in a minimum with little dependence on T \citep{ru03}. In this hypothesis the emission from lines due to highly ionized elements like N~III, N~IV and C~V is allowed.
On the other hand, we can not exclude the possibility of an extreme toplighting effect \citep[concerning a flip of the P-Cygni profile, see][]{top} relative to \Hb\/ and not observed in the \Ha\/ due to a difference in the optical depth (maybe due to the interaction). However to observe this effect the interaction should be really strong while nothing in the light curve suggests such strong interaction. Actually, the comparison with SN 2006bp and SN 1998S shown in panel a) of Fig.~\ref{fig:cfr} suggests the identification of this feature as a blend of highly ionized N and C. Indeed, the feature is closer in wavelength to that observed in SN 1998S, identified by \citet{fa01} as CIII/NIII emission, rather than that in SN 2006bp attributed to He II $\lambda$4686 by \citet{qu07}. Note however that the difference between the FWHM of our feature (140\AA) and the CIII/NIII of SN 1998S ($\sim$20\AA) is remarkable. The similarity with SN 1998S becomes more convincing if one compares a spectrum of 2-3 weeks after the explosion (FWHM of the blended feature $\sim$ 180\AA\/, cfr. Fig.~\ref{fig:cfr}, panel a).

The spectra at 18d and 20d are characterized by a blue continuum, comparable to that of SN 1992H and slightly bluer than that of SN 1999em at a similar age (cfr. Fig.~\ref{fig:cfr} panel b \&~Fig.~\ref{fig:vel}). 
In the optical domain we identified H Balmer lines, He I 5876\AA\/ and some Fe II multiplet lines.
In this set of spectra a narrow \Ha\/ emission due to an underline H II region is visible.

\begin{figure}
\includegraphics[width=8.8cm]{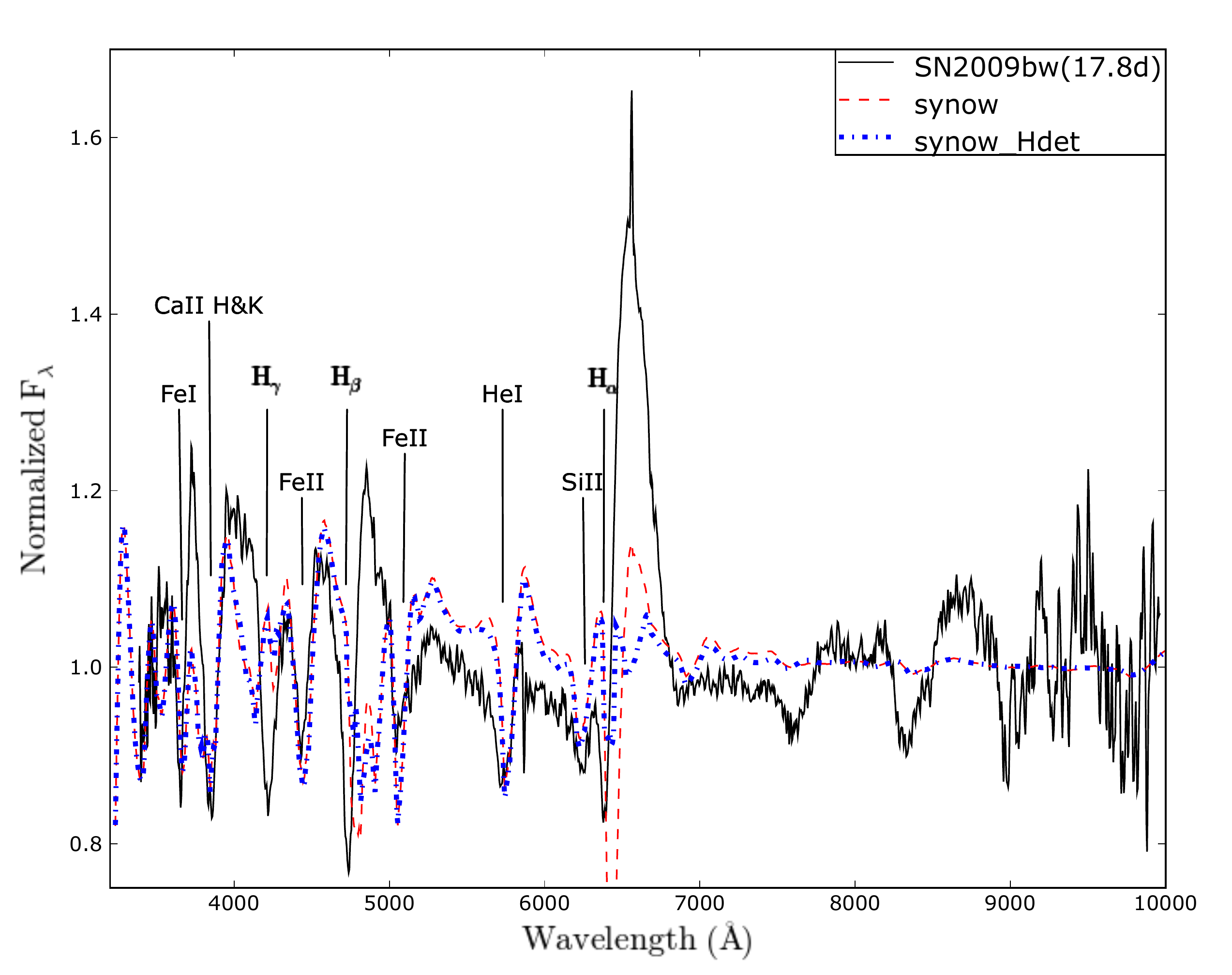}
\caption{Comparison between the optical spectrum of SN 2009bw at 17.8d after the explosion (JD 2454916.5) and SYNOW synthetic spectra (for the composition of the synthetic spectra, see the text). Two models are plotted having H detached (blue dash-dotted line) or undetached (red dashed line). The observed spectrum has been corrected for extinction in the Galaxy, reported to the local restframe and normalized to its continuum. The most prominent absorptions are labelled.} 
\label{fig:id1}
\end{figure}

In order to perform a detailed line identification, we computed a synthetic photospheric spectrum model using SYNOW \citep{sy} and matching the properties of our observed +18d spectrum. We adopted 
a blackbody temperature T$_{bb}\sim$12000 K, an expansion velocity at the photosphere v$_{phot}\sim$7000 \kms\/ (cfr. Sect.~\ref{sec:ev}), an optical depth $\tau(v)$ parameterized as a power law of index $n =$ 9 and  an excitation temperature T$_{exc}$=10000 K, constant for all ions. The most relevant spectral features are reproduced using only 6 ions (Fig.~\ref{fig:id1}). The P-Cygni profiles of the Balmer lines are clearly visible, as well as He I $\lambda$5876, but there is also contribution from Ca II, Fe II, Fe I and Si II. Detached H lines better match the line troughs. The poor fit of the Ca II IR triplet is due to low optical depth, which is however necessary to reproduce the Ca II H \& K lines. We excluded the O I ion because of the early phase of the spectrum.

The line at about 6250\AA\/  has been identified as Si II $\lambda$6355 with an expansion velocity comparable to those of the other metal ions. The presence of Si II 
has been identified in several type II SNe, such as 2007od \citep{07od}, 2005cs \citep{05cs}, 1992H \citep{92h}, 1999em \citep{99em3} and 1990E \citep{90e}.

In the NIR spectrum (20d), having a photospheric temperature derived from a blackbody fit of $\sim$7900 K (cfr.Fig.~\ref{fig:spec_nir}) the P-Cygni profiles of the Paschen series are detected, in particular Paschen $\beta$ and Paschen $\gamma$. The latter is blended with He I $\lambda$10830, and has a velocity which is higher by $\Delta$v$\sim$2200 \kms\/ than that of He I 5876\AA\/.

The subsequent set of spectra (33d-39d) shows well developed P-Cygni profiles of Balmer and metal lines. Sc II $\lambda$5031 on the red side and Sc II $\lambda$4670 on the blue side of \Hb\/ are visible, as well as the Fe II multiplet 42 lines ($\lambda\lambda$  4924, 5018, 5169), Fe II at about 4500\AA\/, Fe I, and Sc II in the region between 5200\AA\/ and 5500 \AA\/. Starting from day $\sim$34, the He I $\lambda$5876 forms a blend with  the Na ID. The Ca II IR triplet ($\lambda\lambda$ 8498, 8542, 8662) is well developed, while the H\&K Ca II feature blends with Ti II. 
O I at about 7700\AA\/ starts to be visible.
Si II $\lambda$6355 has now disappeared, and the bottom of the \Ha\/ absorption shows a flat profile. We will discuss in detail the line profiles in Sect.~\ref{sec:dis}.

\begin{figure*}
\includegraphics[width=18cm]{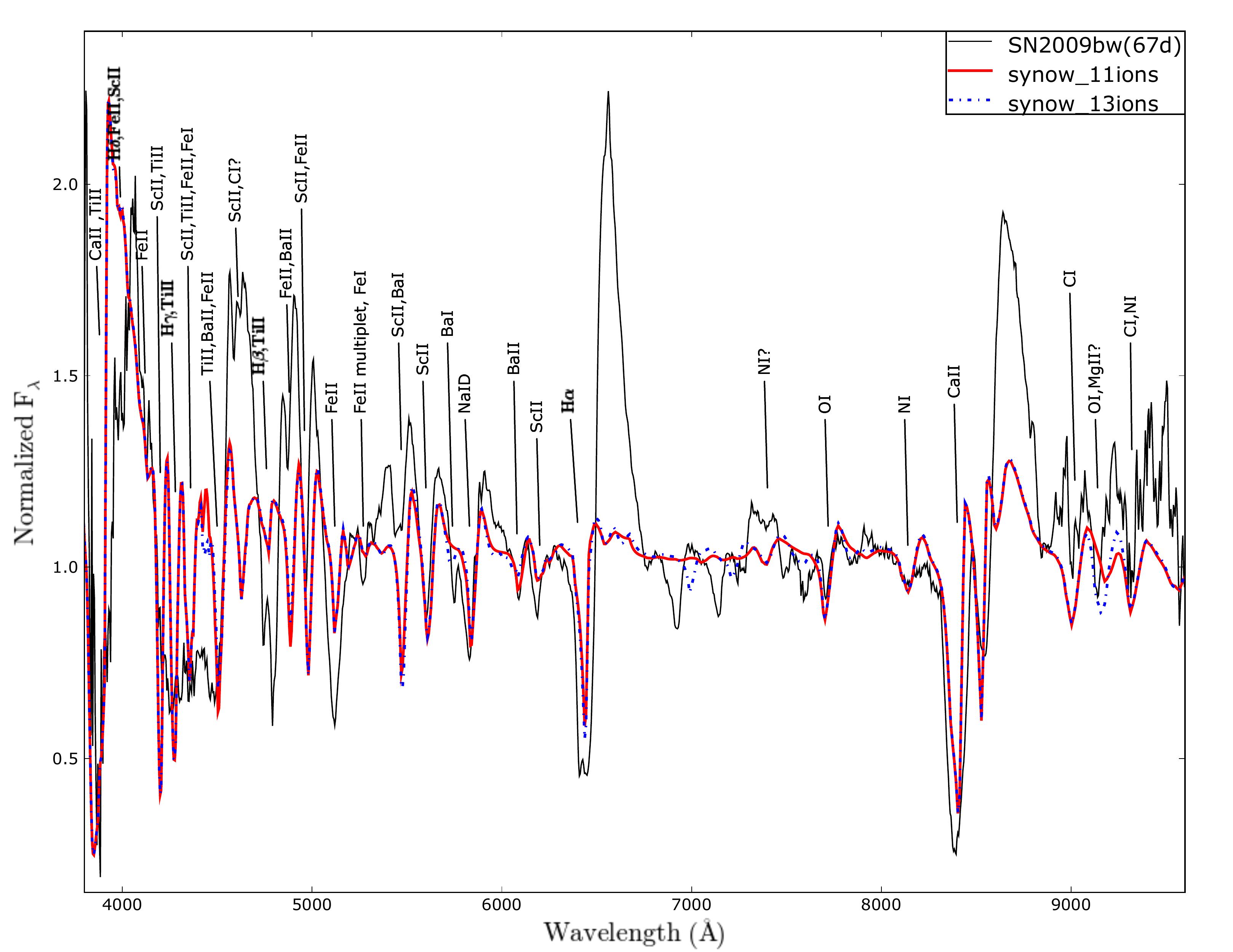}
\caption{Comparison between the optical spectrum of SN 2009bw at $\sim$67d after explosion (JD 2454916.5) and two SYNOW analytical spectra. The difference between the SYNOW spectra is the presence of Ba I and Mg II ions (blue). The spectrum has been corrected for extinction in the Galaxy and reported to the SN restframe. The most prominent absorptions are labelled.} 
\label{fig:id2}
\end{figure*} 

In the late plateau phase the spectra continue to show the elements identified at earlier phases, complemented by Ba II and Ti II, whose presence indicate a temperature of $\sim$5000~K. 
Both ions are required for explaining several features between 4000 and 5000 \AA\/. 
In particular, Ba II $\lambda$6142 is clearly visible. At this phase Ba I may contribute to the blends of lines in the wavelength range 5000--6000\AA\/, and Mg II around 9140\AA\/. 

SYNOW fairly reproduces the characteristics of the best signal-to-noise spectrum obtained in the late plateau phase ($\sim$67d) with the following parameters:
T$_{bb}\sim$5400 K, v$_{phot}\sim$3050 \kms\/ (cfr. Sect.~\ref{sec:ev}), optical depth $\tau(v)$ parameterized as a power law of index $n =$ 7, and T$_{exc}$=5000--8000 K for neutral atoms and 10000 K for ionized species. The SYNOW spectrum (Fig.~\ref{fig:id2}) has been obtained by considering 11 different contributing ions (red dashed line): H I, C I, N I, O I, Na I, Ca II, Sc II, Ti II, Fe I, Fe II, Ba II. The inclusion of two additional ions (blue dash-dotted line), Ba I and Mg II, allows for a better fit of the absorption on the blue wing of Na ID at 5745\AA\/ and the absorption at 9145\AA\/ as a  blend of O I and Mg II. The absorption feature visible around 7100\AA\/ and poorly fitted by both the synthetic spectra could be Ti II with a different optical depth. However, because of the noise and the presence of telluric absorptions in this range it is difficult to draw a firm conclusion. 
The discrepancy in the emission between the SYNOW fits and observed spectra at about 4500 \AA\/ is perhaps due to a high optical depth of Ti II that masks the emission of nearby lines. Other discrepancies between observed and synthetic spectra may be attributed to NLTE effects ignored in our computation.
H I and Ca II are detached to improve the line fit, but this causes problems in fitting the red wings. 
Interesting lines are identified on the red side ($>$7000\AA\/) of the spectrum, where absorptions due to the elements of the CNO cycle appear. In particular, O~I $\lambda$7774 is present, and possibly also O I $\lambda$9260, blended with Mg II. N~I has been identified around 7400\AA\/ and 8130\AA\/. C~I appears around 9010\AA\/ and 9310\AA\/ and possibly around 4615\AA\/ close to the Sc~II line, but in this region the fit is affected by the emission profile of detached hydrogen.
O~I and N~I have been identified at this phase in other SNe, e.g. 1999em \citep{pphdt}, but the presence of C~I  lines in spectra of type IIP SNe has only occasionally been claimed, e.g. in SNe 1995V \citep{fassia} and 1999em \citep{pphdt}.

Fig.~\ref{fig:cfr} (panel c) shows the comparison of spectra of several SNe II in the plateau phase. 
The spectrum of \bw\/ appears similar to that of 
SN 1992H, sharing features commonly present in SNe IIP during the plateau phase. The comparison is interesting with SN~1999em around 9000\AA\/, where SYNOW has identified C I lines. 
Unfortunately the other \bw\/ spectra do not cover this wavelength range. The flat absorption profile of \Ha\/  resembles that observed in SN~1999em. In Sect.~\ref{sec:dis} we will discuss this feature arguing that this is due to weak CSM interaction.

Three early nebular spectra of \bw\/ have been collected between 130d and 190d. A narrow component appears at the host galaxy rest position, confirming that it is due to an underlying H~II region. The broad \Ha\/ emission, attributed to the SN ejecta, is also centered at the rest wavelength and shows a residual absorption. Na~ID and Ca~II IR triplet absorptions are clearly visible. The spectra show also evidence of the forbidden emission of the [Ca~II] $\lambda\lambda$7291, 7324 doublet.
Compared to other SNe IIP (panel d of Fig.~\ref{fig:cfr}) is noticeable the complete absence of [O~I] $\lambda\lambda$6300,6363 even in the latest available SN 2009bw spectrum, and the relatively narrow FWHM of the emission lines (FWHM(\Ha) $\sim$1900 \kms).

\begin{table*}
  \caption{Observed blackbody temperature and expansion velocities in \bw. For the Balmer lines the velocities were measured on the red wing component.}
  \begin{center}
  \begin{tabular}{cclccccc}
  \hline
  \hline
  JD & Phase$^{*}$ & T & v($H_{\alpha}$) & v($H_{\beta}$) & v(He I) & v(Fe II) & v(Sc II)  \\
   +2400000 & (days) & (K) & (\kms)& (\kms)&(\kms) &(\kms) & (\kms) \\
 \hline
  54920.5 & 4.0 & 17250 $\pm$ 1000 & 11200 $\pm$ 2000  & 9180 $\pm$ 1000& 9000 $\pm$ 500& & \\
  54934.3 & 17.8 & 11700 $\pm$ 700 & 8122 $\pm$ 200  & 7760 $\pm$ 120& 7658 $\pm$ 122& 6600 $\pm$ 400 & \\
  54936.3 & 19.8  & 8000 $\pm$ 500& 8067 $\pm$ 80& 7682  $\pm$ 150& 7454  $\pm$ 100& 5970 $\pm$ 400 & 5700 $\pm$ 600\\
  54950.3 & 33.8 & 5200 $\pm$ 400& 7313 $\pm$ 300& 6480 $\pm$ 300& & 4828 $\pm$ 300  & 4560 $\pm$ 200\\
  54953.5 & 37.0 & 5300 $\pm$ 500 & 6680 $\pm$ 200 & 6110 $\pm$ 200 &  & 4138 $\pm$ 100  & 4080 $\pm$ 140\\
  54954.5 & 38.0 & 5200 $\pm$ 400& 6660 $\pm$ 300 & & & & 4080 $\pm$ 140 \\
  54955.3 & 39.0 & 5000 $\pm$ 500 & 6670 $\pm$ 200 & 6109 $\pm$ 130 & & 4130 $\pm$ 100 & 4030 $\pm$ 200 \\
  54968.5 & 52.0 & 5050 $\pm$ 250& 6060 $\pm$ 60 & 5400 $\pm$ 400 & &3480 $\pm$ 100  & 3380 $\pm$ 120 \\
  54980.5 & 64.0 & 5200 $\pm$ 300& 5760 $\pm$ 300 & & & 3097 $\pm$ 80  & 3000 $\pm$ 200\\
  54981.5 & 65.0 & 5250 $\pm$ 250 & 5759 $\pm$ 180& & & 3000 $\pm$ 200\\
  54983.5 & 67.0 & 5400 $\pm$ 300 & 5720 $\pm$ 300& 4400 $\pm$ 230 & &3045 $\pm$ 150 & 2980 $\pm$ 200\\
  54509.5 & 93.0 & 4600 $\pm$ 200& 5165 $\pm$ 150 & 3600 $\pm$ 180 & & 2600 $\pm$ 300 & 2500 $\pm$ 150 \\
  55021.5 & 105.0 & & 4430 $\pm$ 100 && & 1960 $\pm$ 300 & 1910 $\pm$ 110\\
 \hline
\end{tabular}
\begin{flushleft} 
 * with respect to the explosion epoch (JD 2454916.5)
\end{flushleft} 
\end{center}
\label{table:op}
\end{table*}

\subsection{Expansion velocity and temperature}\label{sec:ev}

\begin{figure}
\includegraphics[width=\columnwidth]{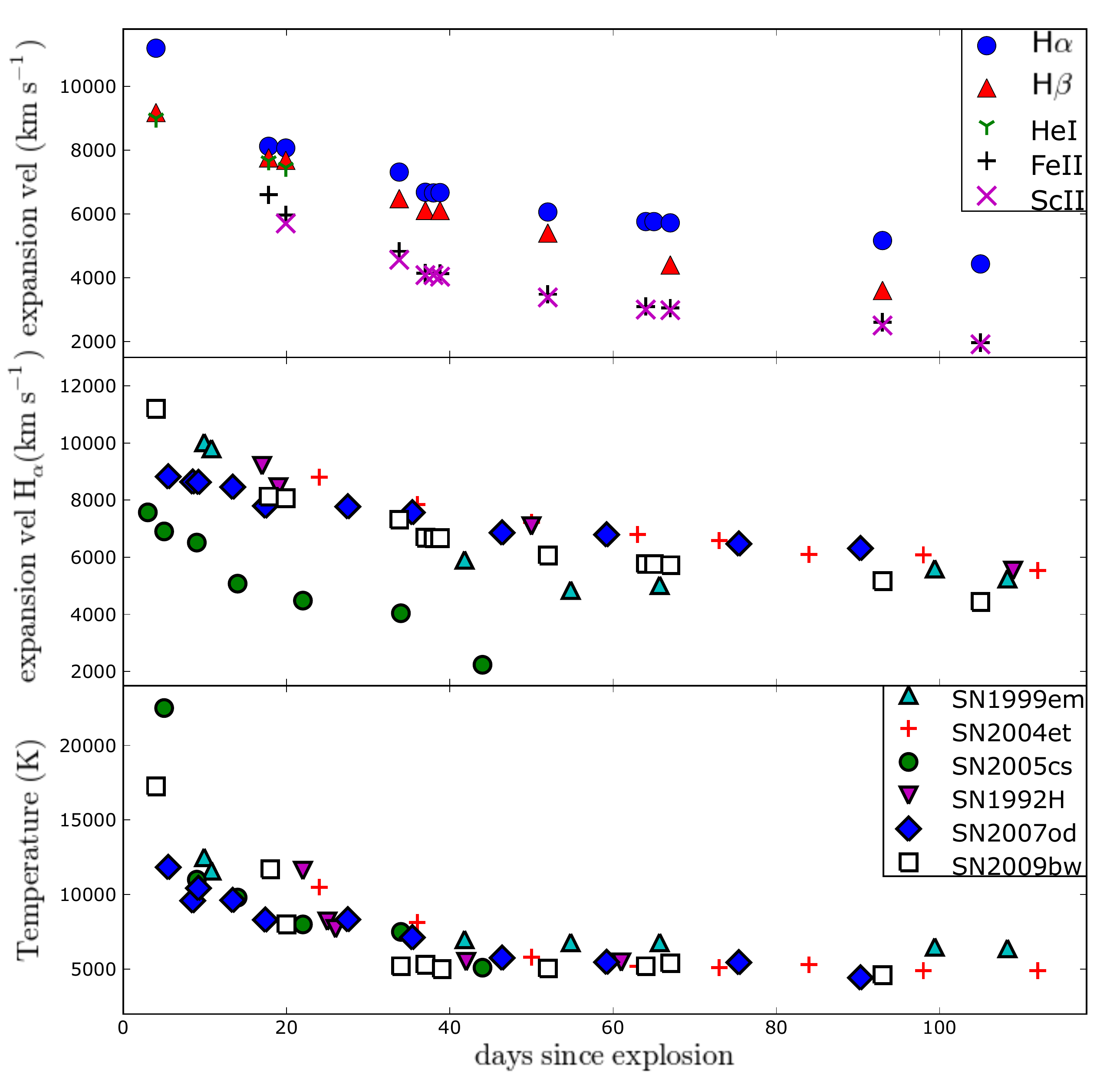}
\caption{Top: expansion velocity of \Ha\/, \Hb\/, HeI $\lambda$5876, FeII $\lambda$5169 and Sc II $\lambda$6246 measured from the minima of P-Cygni profiles. Middle: comparison of the \Ha\/ velocity of SN 2009bw with those of other SNe II. Bottom: Evolution of the continuum temperature of SNe 2009bw, 2007od, 1999em, 2004et, 2005cs, 1992H.}
\label{fig:vel}
\end{figure}

The photospheric expansion velocities of \Ha\/, \Hb\/, He I $\lambda$5876, Fe II $\lambda$5169 and Sc II $\lambda$6246 are reported in Tab.~\ref{table:op} and plotted in Fig.~\ref{fig:vel} (top panel). They have been derived through the minima of P-Cygni profiles. Error estimates have been derived from the scatter of several independent measurements. 
\Ha\/ is always the strongest line, and the derived velocities are systematically the largest.
During the first 20d when He I $\lambda$5876 dominates over Na ID, the He I velocity is comparable with that of \Hb\/.
The Fe II velocity, which is considered to be a good indicator for the photospheric velocity because of the small optical depth, is lower than those of H and He,  reaching 3000 \kms\/ at about two months. Sc II is also a good indicator of the photospheric velocity and indeed its velocity is very close to that of Fe II, supporting the identification of the lines of both ions.

In Fig.~\ref{fig:vel} (middle) we compare the \Ha\/ velocity evolution of \bw\/ with those of other type IIP SNe.
The velocity of \bw\/  in the first two months is close to those of type IIP as SNe 1999em,  2004et, 1992H and 2007od, and higher than that of SN 2005cs which is known to have an exceptionally slow photospheric expansion.
Afterwards, the photospheric velocity of \bw\/ slowly decreases, showing a trend that is slightly different from that of other type II SNe. This may be attributed to the presence of a moderate/low-density CSM that changes the shape of the absorption profile of \Ha\/ (see Sect.~\ref{sec:dis}). 

In Fig.~\ref{fig:vel} (bottom) the evolution of the photospheric temperature, derived from a blackbody fit to the spectral continuum, is shown and is compared to those of the reference sample. We already mentioned that the first spectrum of \bw\/ is quite blue, indicating high black-body temperatures (1.7$\pm$0.1 x 10$^{4}$ K). The temperature remains high also at the second epoch ($\sim$18d). 
Nevertheless, the temperature evolution of \bw\/ is normal and not so different from those of the other SNe in the sample: its ejecta is probably hotter in the first month (similar to \h\/) and marginally cooler than average from the plateau onward. 


\section{Explosion and progenitor parameters}\label{sec:m}

We have inferred the physical properties of the \bw\/ progenitor (namely, the ejected                                                                  
mass, the progenitor radius, the explosion energy) by performing a model/data                                                                        
comparison based on a simultaneous $\chi^{2}$ fit of the main observables (i.e. the bolometric 
light curve, the evolution of line velocities and the continuum temperature at the 
photosphere), using the same procedure adopted for SN 2007od in \citet{07od}.                        

According to this procedure, we employ two codes, the semi-analytic code 
presented in \citet{zamp2} to perform a preparatory                                                                     
study in order to constrain the parameter space, and a new code which includes      
an accurate treatment of the radiative transfer and radiation hydrodynamics                                                                        
\citep[][]{pumo2,puza11}, to narrower grid                                                                     
of models.

The application of these two codes is appropriate if the SN emission 
is dominated by the expanding ejecta. For SN 2009bw the contamination from interaction may in 
part affect the observables (cfr. Sect.~\ref{sec:dis}) and the possibility to reproduce the observed features is to be taken with caution. However, since there is no evidence that the ejecta--CSM interaction is a strong effect during 
most of the post-explosive evolution, our modeling 
can still be applied to SN 2009bw, returning a reliable estimate of the physical parameters of the explosion that can be useful to characterize its progenitor.

The distance and shock breakout epoch are taken from Tab.~\ref{table:main}. 
In order to evaluate the uvoir (from optical to NIR) bolometric luminosity, 
we further assume that the SN has the same SED evolution as SN 1999em and hence
\begin{center}
$L_{09bw} = (L_{99em} / L^{UBVRI}_{99em}) \times L^{UBVRI}_{09bw},$
\end{center}                                                                                                                                             
where $L_{09bw}/L_{99em}$ and $L^{UBVRI}_{09bw}/L^{UBVRI}_{99em}$ are the uvoir bolometric and quasi bolometric
luminosity of SN 2009bw and SN 1999em \citep[][]{99em}, respectively.                  

Assuming a \ni\/ mass of 0.022 \M\/ (see Sect.~\ref{sec:bol}), the best fits of the semi-analytic and 
numerical models are in fair agreement and return values for the total (kinetic plus thermal) energy of 
$\sim 0.3$ foe, an initial radius of $3.6-7 \times 10^{13}$ cm, and an envelope mass of $8.3-12$ \M\/. 

\begin{figure}
 \includegraphics[width=\columnwidth]{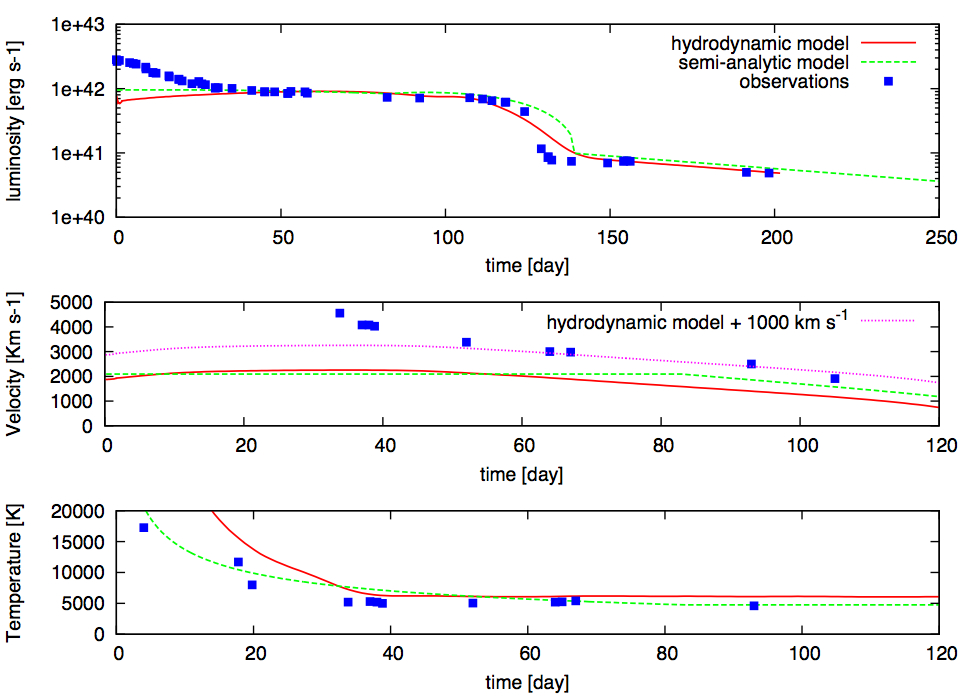}
 \caption{Comparison of the evolution of the main observables of SN
   2009bw with the best-fit models computed with the semi-analytic
   code (total energy $\sim 0.3$ foe, initial radius $3.6 \times
   10^{13}$ cm, envelope mass $8.3$ \M) and with the relativistic,
   radiation-hydrodynamics code (total energy $\sim 0.3$ foe, initial
   radius $7 \times 10^{13}$ cm, envelope mass $12$ \M).  Top,
   middle, and bottom panels show the bolometric light curve, the
   photospheric velocity, and the photospheric temperature as a
   function of time, respectively. To estimate the photospheric
   velocity from the observations, we used the value inferred from the Sc II
   lines (often considered to be a good tracer of the
   photosphere velocity in type II SNe). The dotted-purple line 
   is the radiation-hydrodynamics model shifted by 1000 \kms\/ to reproduce the later epochs of data (see text).}
 \label{fig:model}
\end{figure}

The agreement between our modeling and the observed luminosity and temperature evolution is reasonably good, except at early epochs ($\lesssim$ 33d, cfr. Fig. \ref{fig:model}).
These discrepancies may in part be caused by some ejecta-CSM interaction (see discussion in Sect.~\ref{sec:dis}), 
leading to an excess of the observed luminosity, and to the approximate initial density profile 
used in our simulations, that does not reproduce correctly the radial profile of the outermost 
high-velocity (HV) shell of the ejecta formed after shock breakout \citep[cfr.][]{puza11}

During the entire photospheric phase the best fit of the velocity is unsatisfactory. This is unusual for the Pumo and Zampieri code that has proved to work well for standard SNe II, reinforcing the idea that for \bw\/ the ejecta--CSM interaction could be at work. Indeed the discrepancies between the simulated photospheric velocity and the data could be related to the interaction creating of a pseudo-photosphere at a larger radius, hence at a higher velocity ($\Delta v\sim 1000$ \kms\/), than expected by the model (cfr. Fig. \ref{fig:model}, middle panel, purple-dotted line). This effect has been found in luminous and interacting SNe as shown by \citet{agn09}.

The model parameters reported above are consistent with both a super-asymptotic giant branch (SAGB) star, with a mass close to the upper limit of this class, and a progenitor of a small Fe core collapse SN with a mass close to the lower limit. They are reliable progenitors even if, through the first one is difficult to explain the amount of \ni\/ \citep[e.g.][]{wa09}, instead the second one poorly explains the mass loss and the probable mixing related to the CNO elements \citep[e.g.][]{ma08}. However, in both scenarios the progenitors are expected to have a radius of $\sim$1000 R$_{\odot}$ \citep{gar94,rit99,woo02} and hence to fall into the red supergiant (RSG) category (200 R$_{\odot}<$ R $<$ 1500 R$_{\odot}$).

\section{Discussion}\label{sec:dis}
In the previous Sections we have presented and discussed the photometric and spectroscopic data of  \bw\/ in UGC~2890 from the photospheric to the early nebular stages.

The analysis reported in Sect.~\ref{sec:bol} indicates that \bw\/ was a relatively luminous SN IIP with a peak magnitude similar to that of SN 1992H (M$_{V}=-17.67$) and an extended plateau in V, R and I.
The duration of the plateau suggests an envelope mass not dissimilar from that of standard SNe IIP such as SN 1999em, that in the nebular phase shows strong similarity with \bw\/ both in photometry and spectroscopy. 
The rise to maximum seems relatively slow for a SN IIP ($\sim8$d) and the peak somehow broad.
The light curves show a steep but shallow drop of $\sim$2.2 mag in $\sim$13d from the end of the plateau to the tail. The tail is 
flat for the first couple of months, then declines with the canonical $^{56}$Co decline rate of 0.98 mag/100d.
The luminosity of the radioactive tail indicates an ejected mass of $^{56}$Ni of M(\ni)$\sim$ 2.2x10$^{-2}$ M$_{\odot}$ (cfr. Tab.~\ref{table:main}), 
which is comparable to that estimated for SN 1999em \citep{99em}. 

The spectral evolution is quite similar to that observed in canonical SNe IIP, although an unidentified, broad feature is present 
in the first photospheric spectrum, at about 4600\AA. A plausible interpretation is that the feature is due to emission emerging in a region of low opacity. 
 
The host galaxy has been classified as an highly inclined peculiar Magellanic Spiral (Sdm pec:). 
In the spectra an unresolved component due to an underlying H~II region is clearly visible.
A reasonable assumption is to consider the SN has the same metallicity of the H~II region.
From the spectra obtained at Calar Alto on Sept. 4 and at TNG on Oct. 11, 2009, we have measured the N2 and (only for the CAHA spectrum) the O3N2 indices \citep{pp04} of the H~II region extracted close to the SN along the slit of the spectrograph. The average relations of \citet{pp04} then provide the O abundances which turn out to be 12+log(O/H) = $8.66\pm0.06\pm0.41$ (where the first error is statistical and the second one is the 95\% spread of the N2 index calibration relation).
While the projected distance from the galaxy nucleus is relatively small (2.4 kpc) we were unable to determine the deprojected one because of the very high inclination  (i$=90\degr$) of the parent galaxy (Sect.~\ref{sec:intro}). 
\citet{pily04} have shown that the central oxygen abundances of spiral galaxies is typically 8.3 $<$ [12+logO/H] $<$ 9.1, similar to solar abundance \citep[8.69,][]{asp09}.
Therefore, the estimated progenitor metallicity is close to solar and in line with the expected metallicity at the position inside the parent galaxy.

No information is available in the literature about the X-ray and radio emission of the SN.
An observation (8 ksec) with Swift-XRT has been obtained generating a single combined and astrometrically-corrected event file. 
In an aperture of 9 XRT pixels ($\sim$21") centered at the position of \bw\/ we obtained a limit on total source counts of n~$<$~5. Considering a  column density of N$_{H}$=~1.9~$\times$~10$^{21}$~cm$^{-2}$ \citep{dilo90,pre95}
we have measured an upper limit to the X-ray flux (over the energy range 0.3--10 keV)
$7.9\times10^{-14}$ erg cm$^{-2}$ s$^{-1}$ corresponding to upper limits of the X-ray luminosity of L$_X < 3.77 \times 10^{39}$ erg s$^{-1}$ (thermal bremsstrahlung model with kT = 10 kev) and $1.6\times10^{-13}$ erg cm$^{-2}$ s$^{-1}$ corresponding to  L$_X < 7.65 \times 10^{39}$ erg s$^{-1}$ (power law model with photon index $\Gamma$=1.7).
These X-ray luminosity upper limits do not preclude the possibility of a weak interaction as in SN IIP 1999em.
Indeed, SN~1999em was detected at a similar phase at fainter flux ($\sim 10^{-14}$ erg cm$^{-2}$ s$^{-1}$ in both the 2-8 keV and 0.4--2 keV bands) thanks to the superior performance of Chandra.
The corresponding X-ray luminosity is one order of magnitude smaller than our upper limit \citep[L$_X$(99em) $\sim$ 2 $\times$ 10$^{38}$  erg s$^{-1}$, in the 0.4-8 keV range,][]{pooley} because of the much shorter distance.


\begin{figure*}
\includegraphics[width=13 cm]{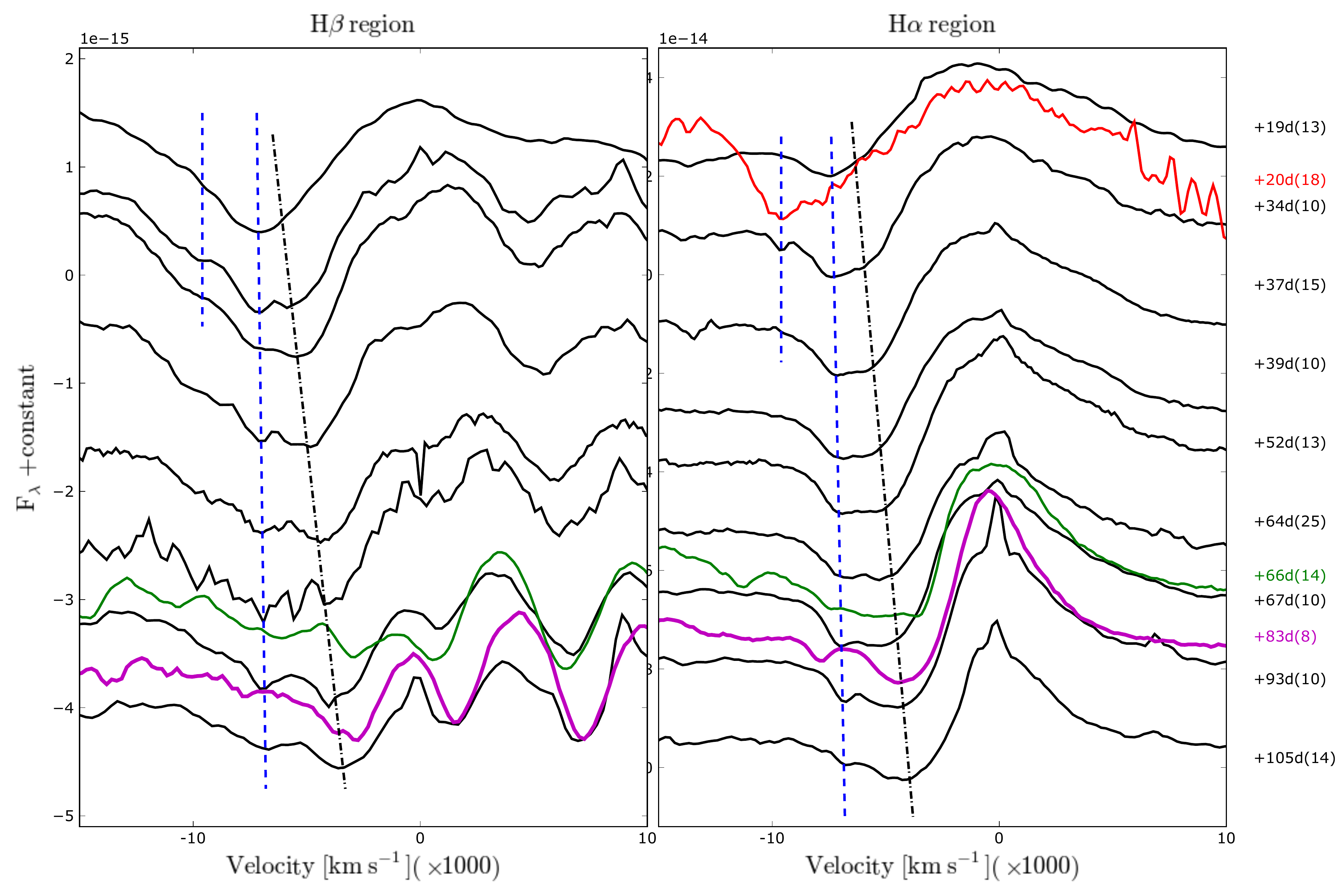}
\caption{Zoom of the \Hb\/ (left-hand panel) and \Ha\/ (right-hand panel) spectral regions during the plateau phase of \bw\/. The {\it x}-axes are in expansion velocity coordinates with respect to the rest-frame positions of the lines. In order to guide the eye, two black dash-dotted lines are drawn in the spectra corresponding expansion velocities. Similarly two blue dashed lines follow the HV features of the Balmer lines. The red spectrum (+20d) shows the NIR He I feature centered at 10830\AA. We have also reported the spectra of SN 2004dj (purple, +83d) and SN 1999em (green, +66d) for comparison.}
\label{fig:hv}
\end{figure*}

However, some observational evidence point in favour of weak CSM interaction in \bw\/.
The spectral line contrast in the early photospheric period seems smaller than in other SNe IIP (cfr. panel a \& b of Fig.~\ref{fig:cfr}), as expected in the case of resonance scattering due to external illumination of the line forming region by CSM interaction \citep[the so called toplighting,][]{top}. 
In addition, the spectra show the presence of secondary absorptions in the H Balmer lines that we interpret as HV features.
The right panel of Fig.~\ref{fig:hv} shows the presence of a blue  \Ha\/ component at about 7300 \kms\/  that does not evolve with time, and a redder component decreasing progressively in velocity in analogy with other photospheric lines measured in Fig.~\ref{fig:vel}.
Careful analysis of \Hb\/ (left panel of Fig.~\ref{fig:hv}) shows the same bluer non-evolving component as well as the photospheric one, despite the optical depth of the line being smaller than that of \Ha.
Blue secondary features, constant with time, were also identified in SNe~1999em and 2004dj by \citet{chugai}.
According to these authors, the lines are HV components of the Balmer lines. Their analysis shows that the interaction of the ejecta of a SN IIP with an average RSG wind can be detected during the photospheric stage through the emergence of absorptions (shoulders) on the blue wings of the undisturbed H lines due to enhanced excitation of the outer unshocked ejecta. In \bw\/ the unshocked layers producing the HV absorptions are at about 7300 \kms\/, while the photospheric absorption moves from 6700 to 4400 \kms\/ between the early and the very late photospheric phase (day 37 to 105).
\citet{chugai} also predicted that during the late photospheric phase the physical conditions in the cool dense shell (CDS) behind the reverse shock allow the formation of a notch, a smaller and narrower absorption, at the fastest edge of the line, and recovered it in the spectra of SN~2004dj. We clearly see this notch with similar intensity and position in \bw\/ starting from 2 months past explosion, both for \Ha\/ and \Hb\/.
The line position is compatible with the expected position of Fe II lines at the earliest epoch, but the fact that it does not evolve disfavors such an identification.

Only one NIR spectrum is available, at an epoch (21d) when the two components of the Balmer lines are blended.
The NIR spectrum shows a strong feature at about 10500\AA\/. If  we identify it as He I $\lambda$10830, the expansion velocity turns out  to be larger (v$_{He}\sim10000$ \kms) than that of He I $\lambda$5876 (which has no sign of a photospheric component), and larger than that of
the HV Balmer components.
A NIR spectrum of SN~1999em at a similar epoch is available \citep{99em}, showing an absorption identified by \citet{chugai} as HV He I at a velocity 
which was marginally higher than those of H and without a photospheric component.
The alternative identification with Mg II $\lambda$10926 or Fe II $\lambda$10862  would imply even larger velocities.
In conclusion, the close similarity with the features observed in \em\/ and SN 2004dj and well modeled by \citet{chugai}, leads us to favour
the HV Balmer component scenario.
Line profile models suggest a typical red supergiant wind with a density of $w=\dot{M}_{-6}/u_{10}\sim1$, i.e.
$\dot{M}\sim10^{-6}$ M$_{\odot}$ yr$^{-1}$, similar to those of SNe 1999em and 2004dj \citep[][]{chugai}, or even higher. 
Assuming a typical duration of $10^6$ yr, the mass lost by the progenitor star is about 1M$_{\odot}$.


Late photospheric spectra of \bw\/  show absorption features of CNO elements (see Sect.~\ref{sec:sa}), indicating a potentially enhanced mixing with deep layers, which may be an indicators of a high metallicity envelope, prone to enhanced mass loss. 
Similar features in the red part of the optical spectra, related to the CNO elements, have been seen also in SN 1999em \citep{pphdt} and SN 1995V \citep{fassia}, both presenting signatures of weak CSM interaction \citep{pooley,chugai}. 
The presence of C in the photospheric spectra rises the possibility of CO molecules formation, that with their rotation-vibration states are a powerful coolant and are a necessary condition for dust condensation in the ejecta. 
Indeed SN~1999em showed both C in the photospheric spectra and dust formation in the inner ejecta after t$\sim$500d \citep{99em}. 
In \bw\/ there is no direct evidence of dust formation (neither photometric nor spectroscopic) though the last two points of the light curves, affected by large uncertainties, may indicate a steepening of the decline with respect to the slope of $^{56}$Co.

\citet{moriya} studied the interaction between SN ejecta and the CSM around RSGs. They showed that, if the temperature and the 
CSM density are high enough, the CSM around SNe IIP becomes optically thick and the effective photosphere forms at large radii inside the CSM.
The interaction-powered phase is characterized by light curves with broad peaks, flat plateau brighter for large $\dot{M}$ and longer for an extended CSM.   
The observed expansion velocity at early times is expected to be that of the CSM, then turning to typical SN values when the photosphere recedes into the ejecta.
In particular, the early light curves of very bright SNe IIP can be interpreted in terms of interaction with a dense CSM produced by a mass loss at a rate larger than $\sim10^{-4}$ M$_{\odot}$ yr$^{-1}$.
The interaction ejecta--CSM applied to the luminous SN~2009kf \citep{bott10} satisfactorily explained the early multicolor light curves, but failed in explaining the late behaviour and the kinematics, possibly because it was an energetic explosion with large \ni\/ production.
We have seen above that the line profiles of \bw\/  would point toward a normal RSG wind with a low value of $\dot{M}$, which is of the order of $\sim$10$^{-6}$ M$_{\odot}$ yr$^{-1}$ (or marginally higher). This is much smaller than the values expected to significantly increase the luminosity. In addition, we have not detected any unusual behaviour in the velocity evolution.
Nevertheless, we cannot exclude that a fraction of the observed luminosity during the plateau of \bw\/ might be due to the transformation of ejecta kinetic energy into radiation via interaction with the CSM. 
As already highlighted in Sect.~\ref{sec:m}, this may explain why, differently from our previous experience \citep{puza11,07od,09E} where the main observables are simultaneously reproduced by our modelling, here we fail to reproduce the kinetic evolution, whilst
both the bolometric light curve (cfr. Fig.~\ref{fig:cfr_bol}) and the spectra in the nebular phase (cfr. panel d of Fig.~\ref{fig:cfr}) are standard, and very similar to those of SN 1999em.

As reported in Sect.~\ref{sec:m}, the ejecta mass is $\sim8-12$ M$_{\odot}$. 
Accounting for a compact remnant (NS) of $\sim$ 1.6 M$_{\odot}$ and about 1 M$_{\odot}$ of mass loss, the initial mass of the progenitor is of the order of $11-15$ M$_{\odot}$, comparable with that of massive SAGB stars or a low mass Fe CC-SN progenitor. Despite that, we want to notice that the weak intensity of [O I] $\lambda\lambda6300-6363$ may be inconsistent with the progenitor mass (too high for the intensity observed), but the similarity between the late spectra of \bw\/ and \em\/ and the previous modelling case of the weak interacting \od\/ \citep{07od}, are in favour of this mass range.


\section{Conclusions}\label{sec:final}

In this paper we show, for the first time, photospheric and spectroscopic data of \bw\/. The peak (M$_{R}= -17.82$) and the plateau (M$_{R}= -17.37$) magnitudes allow us to accommodate this SN in the small sample of the brightest type IIP SNe. The fast jump from the plateau phase to the nebular tail ($\sim$2.2 mag in 13d, unusual among  bright type II SNe), the proof of the presence of CNO elements in the photospheric spectra and the detection of HV features in the Balmer lines during the recombination phase suggesting an early interaction, make this object atypical in the context of type IIP events.

The well sampled light curve of \bw\/ reaches a luminous plateau at M$_{R}= -17.37$ between $\sim$50d and $\sim$100d. This is a standard duration for the plateau phase in SNe IIP. The high photospheric luminosity contrasts with the moderate luminosity of the tail (similar to that of \em\/). The post-plateau decline is relatively rapid, resembles that of SN 1986L \citep[although this object has not been well studied;][]{hphdt} and has been well monitored, making this constraint rather solid. The remarkable magnitude drop is analogous to that observed in faint SNe IIP \citep{pa1}. The bolometric light curve tail follows the slope expected if the main energy source is the decay of \co\/ into \fe\/. The \ni\/ mass derived from the quasi bolometric (U to I) light curve is M(\ni)$\sim$0.022 \M\/, similar to that of the prototypical SN IIP 1999em. 
If instead, the \ni\/ mass is estimated from the post plateau drop with a steepness {\it S}~=~0.57, M(\ni\/)$\sim$0.002 \M\/ is obtained. This value suggests that for \bw\/ the anti-correlation between the steepness function and $^{56}$Ni mass does not work 
in this case, possibly masked by the presence of other effects, such as CSM interaction or dust formation.


Thanks to detailed synthetical spectral computed using the LTE code \texttt{SYNOW}, we noted the presence of the Si II $\lambda$6355 feature in the early spectra, and of CNO element lines in the plateau spectra. Highly ionized C and N features have been tentatively identified in our first spectrum as being responsible for the prominent blend at 4600\AA\/.
HV line components of the Balmer series (\Ha\/ and \Hb\/ at v$\sim$7300 \kms\/) have been clearly identified in the plateau spectra from $\sim$37d onwards, though they become more prominent from day 52 to day 105. These lines, similar to those observed in SNe 2004dj and 1999em \citep{chugai}, suggest an early interaction with a barely dense CSM. Based on the similitudes between \bw\/ and \em\/, the absence of remarkable changes in the light curve \citep[see][]{moriya} and the early line velocity evolution, we roughly estimate the progenitor's mass loss to be in the range $10^{-6}<\dot{M}<10^{-4}$ \M\/ yr$^{-1}$, even though it is probably closer to the lower limit. Assuming a typical duration of $10^6$ yr, the mass lost by the progenitor star is 1M$_{\odot}$.
Our latest spectra have been useful to estimate the N2 and O3N2 indices \citep{pp04}, revealing an average oxygen abundance of 12+log(O/H)$\sim$8.66, which corresponds to about solar metallicity. 

The presence of CNO elements, the similarities with \em\/ at late phase and the increased slope of the radioactive tail as suggested by the last two photometric points (although with a large uncertainty) might indicate
late dust formation in the inner ejecta.

Modelling gives to us an ejecta mass of $8-12$ \M\/, corresponding to an initial mass of the progenitor of the order of $11-15$ \M\/. This is consistent with a RSG, which could be both with a massive SAGB star or a star exploded as small Fe CC-SN, explaining the most part of the peculiarities (early interaction, \ni\/ mass and CNO elements) of this object.
The unsatisfactory fit of the velocity during the entire photospheric phase strengthens the idea of ejecta--CSM interaction. In fact, the interaction could have caused the formation of a pseudo-photosphere at a radius larger than expected by the model.

\bw\/ shows some properties in common with both luminous and standard SNe IIP.
Our study has revealed: - HV features as in SNe 1999em, 2004dj and 2007od; - ejecta-CSM interaction ongoing during the entire photospheric phase, reinforcing the idea that interaction can be significant also for SNe IIP; - and the presence of CNO elements seen in few SNe IIP. 
Luminous SNe IIP with weak CSM interaction may bridge the gap between normal SNe IIP and interaction dominated SNe IIn, and may thus be helpful to create a united scheme of all CC explosions.

\section*{Acknowledgments}
C.I., S.B., F.B., E.C. and M.T. are partially supported by the PRIN-INAF 2009 with the project "Supernovae Variety and Nucleosynthesis Yields" and by the grant ASI-INAF I/009/10/0.
EB  was supported in part by SFB 676, GRK 1354 from the DFG, NSF grant AST-0707704, and US DOE Grant DE-FG02-07ER41517.
M.L.P. acknowledges the financial support by the Bonino-Pulejo Foundation. The TriGrid VL project, the {\it ``consorzio COMETA''}  and the INAF - Padua Astronomical Observatory are also acknowledged for computer facilities.
S.T. acknowledges support by the  Transregional Collaborative Research Cebter TRR33 \textquotedblleft The Dark Universe" of the German Research Fondation (DFG).
D.Y.T. and N.N.P. were partly supported by
the Grant 10-02-00249 from RFBR.
We thank the support astronomers at the Telescopio Nazionale Galileo, the Copernico Telescope, the 2.2m Telescope at Calar Alto, the Liverpool Telescope, the Nordic Optical Telescope, the SAO-RAS Observatory and the Taurus Hill Observatory for performing the follow-up observations of \bw. 
C.I. thanks the Oklahoma University for the hospitality.
This research has made use of the NASA/IPAC Extragalactic Database (NED) which is operated by the Jet Propulsion Laboratory, California Institute of Technology, under contract with the National Aeronautics and Space Administration. 
We acknowledge the usage of the HyperLeda database (http://leda.univ-lyon1.fr). We also thank the High Energy Astrophysics Science Archive Research Center (HEASARC), provided by NASA's Goddard Space Flight Center, for the SWIFT data.
The Supernova Spectrum Archive (SUSPECT, $http://suspect.nhn.ou.edu/~suspect/$) were used to access data and the availability of this service is gratefully acknowledge.

\label{lastpage}

\end{document}